\documentclass[10pt,twocolumn]{article}

\usepackage{graphicx}
\usepackage{multicol}
\usepackage{amssymb}
\usepackage{amsmath}
\usepackage{booktabs,multirow}
\usepackage{color}
\usepackage{url}
\usepackage{cite}
\usepackage{scrextend}

\newcommand{\tA}{\mathcal{A}}

\newcommand{\tD}{\mathcal{D}}

\newcommand{\di}{\mathrm{diag}}
\newcommand{\tG}{\mathcal{G}}

\newcommand{\tL}{\mathcal{L}}

\newcommand{\tN}{\mathcal{N}}

\newcommand{\bT}{\mathbb{T}}
\newcommand{\bR}{\mathbb{R}}

\newtheorem{thm}{Theorem}
\newtheorem{rem}{Remark}

\DeclareMathOperator*{\argmin}{argmin}

\date{}

\begin{document}

\title{Optimal Solution Analysis and Decentralized Mechanisms for Peer-to-Peer Energy Markets}

\author{Dinh Hoa Nguyen
\thanks{Dinh Hoa Nguyen is with the International Institute for Carbon-Neutral Energy Research (WPI-I$^2$CNER), and the Institute of Mathematics for Industry (IMI), Kyushu University, Fukuoka 819-0395, Japan. E-mail: $hoa.nd@i2cner.kyushu-u.ac.jp$. }
}

\maketitle

\begin{abstract}
This paper studies the optimal clearing problem for prosumers in peer-to-peer (P2P) energy markets. It is proved that if no trade weights are enforced and the communication structure between successfully traded peers is connected, then the optimal clearing price and total traded powers in P2P market are the same with that in the pool-based market. However, if such communication structure is unconnected, then the P2P market is clustered into smaller P2P markets. If the trade weights are imposed, then the derived P2P market solutions can be significantly changed. Next, a novel decentralized optimization approach is proposed to derive a trading mechanism for P2P markets, based on the alternating direction method of multipliers (ADMM) which naturally fits into the bidirectional trading in P2P energy systems and converges reasonably fast. Analytical formulas of variable updates reveal insightful relations for each pair of prosumers on their individually traded prices and powers with their total traded powers. Further, based on those formulas, decentralized learning schemes for tuning parameters of prosumers cost functions are proposed to attain successful trading with total traded power amount as desired. Case studies on a synthetic system and the IEEE European Low Voltage Test Feeder are then carried out to verify the proposed approaches.    
\end{abstract}


\section*{Nomenclature}
\begin{labeling}{$P_{i,G}$, $P_{l,RDP}$}
\item[MAS] Multi-agent system.
\item[ICT] Information and communication technology.
\item[P2P] Peer to peer.
%\item[DES] Dwelling electricity system.
\item[DER] Distributed energy resource.
\item[ADMM] Alternating direction method of multipliers. 
%\item[DCJ--ADMM] Distributed consensus-based Jacobian alternating direction method of multipliers. 
%\item[ISO] Independent system operator. 
%\item[PV] Photovoltaic.
%\item[FC] Fuel cell.
%\item[CHP] Combined heat and power.
\item [$P_{ij}$, $P_{i}$, $P_{i,tr}$] Traded power between peers $i$ and $j$, vector of peer $i$ traded powers, and peer $i$ total traded power [kW].
\item [$P_{i,tr}^{\min}$, $P_{i,tr}^{\max}$] Lower and upper bounds of total traded amount of peer $i$ [kW].
\item[$\tG, \, \tA, \, \tD, \, \tL$] P2P interconnection graph, its adjacency, degree, and Laplacian matrices.
%\item[$\tE$] Edge set of P2P trading graph.
%\item[$\lambda_{\max}(A)$] Largest eigenvalue of a symmetric matrix $A$.
%\item[$\delta_{ij}, \, \delta$] Communication uncertainties and bound of uncertainty. 
\item[$\mathbf{1}_n$, $I_n$] Vector with $n$ elements equal to $1$, and $n\times n$ identity matrix.
\item[$\di\{\}$, $vec()$] Diagonal or block-diagonal matrices, and stacked vector. 
\item[$\bR$, $\bR^{n}$, $\bR^{n\times m}$] Set of real numbers, real $n$-dimension vectors, and real matrices with dimensions $n\times m$.
\end{labeling}

%%%%%%%%%%%%%%%%%%%%%%%%%%%%%%%%%%%%%%%%%%%%%%%%%%%%%%%%%%%%%%%%%%%%%%%%%%%%%%%%%%%%%%%%%%%%%%%%%%%%%%%%%%%%%%%%%%%%
\section{Introduction}

P2P energy system has recently emerged as an attracting concept for novel energy market designs to push the flexibility, diversity, locality, and low emission of energy supply and consumption, due to the increasing penetration of DERs into energy grids \cite{Baez-Gonzalez18,Sousa19,Tushar18,Tushar20}. First, energy losses are reduced in P2P systems because energy is usually exchanged within short distances. As a result, investment cost could be lower. Second, P2P trading platforms are often decentralized and localized, which are very suitable for integrating DERs and give much more flexibility for prosumers to handle their energy balance and profit. Next, equipped with advanced ICT technologies, e.g. distributed ledger technologies and block-chain, the security and privacy in P2P markets are much better than that in the conventional bulk energy grids \cite{Tushar18}. Last but not least, P2P trading helps promote new businesses since different models and market scales can be performed under P2P platforms, e.g. federated plans \cite{MorstynP2P18}, full P2P, community-based, or their hybrid combination \cite{Sousa19,Moret19}. To this end, P2P energy system will serve as an important base to transform the current top-down, centralized energy networks into bottom-up, decentralized ones. 

Each peer in P2P energy systems is a prosumer who can act as a producer at one time and as a consumer at another time. The trading of prosumers is affected by their preferences, e.g. energy transfer distance, or the sources of generated powers, etc. Usually, successful energy transactions are derived with short transfer distances and clean energy sources to avoid energy losses and reduce emissions of pollutants. Moreover, the role of each prosumer as a buyer (consumer) or seller (producer) can change from a time step to another, because its generation or consumption profile is time-varying and is only known by prediction in advance. %Accordingly, the system structure in P2P energy markets is time-varying. 
 Hence, the optimization problems arising from the optimal energy management or optimal planning and operation of P2P energy systems are essentially different from that in pool-based energy markets. 

Another feature distinguishing P2P energy market with other energy markets is on the direct energy trading between each prosumer (peer) with another communicated prosumer (peer). Therefore, a power balance constraint is forced to each pair of communicated prosumers, instead of only one balance constraint for the total generated and consumed powers of all producers and consumers (e.g. in pool-based markets). In order to deal with this pairing constraint, a number of different P2P trading schemes  has been proposed, e.g. bilateral contracts \cite{Sorin19,Baroche19,MorstynP2P19b,Khorasany19}, game theory based \cite{Tushar19,Tushar18,Moret19,Cadre20}, distribution optimal power flow \cite{GuerreroA19}, supply-demand ratio based pricing \cite{NLiu17}, mixed performance indexes \cite{Werth18}, Lyapunov optimization \cite{NLiu18}, multi-class energy management \cite{MorstynP2P19a}, continuous double auction \cite{GuerreroA19}, etc. 

To solve the optimal energy management problems arising in P2P energy systems,  decentralized optimization approaches are preferred to centralized or  distributed optimization methods, because of the following reasons. First, decentralized approaches naturally fit into the structure of P2P energy systems, where no central coordination unit is needed and each peer directly communicates with other peers. Second, system robustness with respect to failure of individual parts is better with decentralized optimization approaches, because if some parts malfunctions, then the remaining still works, while centralized or distributed optimization methods have a single point of failure which stops the whole system from working if the central unit fails. Third, communication load is burdened at the central unit in centralized methods, but is much lessen at each node in  decentralized approaches. As such, most of the existing optimization algorithms for P2P energy systems hitherto are decentralized.

ADMM is originally a distributed optimization approach which has been most often used in P2P energy systems research \cite{KZhang20,Baroche19,Baroche19conf,MorstynP2P19a}. A direct application of ADMM to allocate exogenous costs in P2P energy systems was presented in \cite{Baroche19,Baroche19conf}, however local ADMM variables at each prosumer need to be updated in order. A distributed ADMM method together with model predictive control were introduced in \cite{MorstynP2P19a} for solving a multi-class energy management problem, but a P2P platform agent exists for calculating dual prices and solving an optimal power flow problem. Coordination of P2P energy trade and ancillary services was investigated in \cite{KZhang20}, where global variables on energy transfer matrix was required in the utilized ADMM algorithm. %The Jacobian Proximal ADMM (JP-ADMM) proposed in \cite{Deng2017}, and a gossiping push-sum algorithm for JP-ADMM were applied in \cite{Almasalma19} for P2P integrated voltage control of photovoltaic inverters. 
The effects of inter-peer communication sparsity to an ADMM algorithm convergence speed and market outcomes and the comparison between P2P market and other markets were studied in \cite{Baroche19conf}. However, \cite{Baroche19conf} considered the full communication structure which incurs more communication loads in P2P energy systems. 

Comparison of different optimization approaches including ADMM with a central coordinator, primal-dual algorithms, and Relaxed Consensus + Innovation (RCI) \cite{Sorin19} for P2P energy systems has been conducted in \cite{Khorasany19}. It turned out that the first method is fastest, while the last method is slowest, which concurs with the observation of RCI slow convergence in \cite{Baroche19conf}.  Note that the existing ADMM algorithms for P2P energy systems require either a central coordinator (i.e. not decentralized), or a sequential update of local ADMM variables at each prosumer though all prosumers work in parallel (potentially longer computational time). 

In all existing studies on P2P energy systems, two fundamental issues are commonly assumed, one is the successful trading of all peers, and the other is the right selection of cost function parameters by each prosumer. However, the first assumption can be violated in realistic situations because of: (i) distinct expectations between prosumers on the amount of powers and their prices to be traded; (ii) different energy preferences from prosumers. Such violation results in completely different solutions with those when the assumption is satisfied, as will be shown later in this paper. The second assumption is also difficult to be satisfied in reality, because prosumers cannot know exactly the values of their cost function parameters for obtaining their expected power trading amount and price.

This paper aims to fulfill the above research gaps, thereby contributes the following to the P2P energy systems research. 
%studies the relation of their optimal solution to that of traditional energy systems, i.e. without P2P trading. 
%Accordingly, the contributions of the current research are summarized in the following.
\begin{itemize} 
	\item A decentralized, scalable ADMM approach with parallel updates of prosumers/peers, which also updates local ADMM variables at each prosumer/agent in parallel, for trading in P2P energy market. Thus, it is suitable for distributed and parallel computing platforms, and its computational time is shorter. 
	\item Decentralized learning strategies for tuning parameters of prosumers cost functions to obtain successful trading with expected amount of total traded power.  
\end{itemize}

The rest of this paper is organized as follows. Section \ref{sysmod} introduces the optimal energy management problem in P2P systems and characterizes its optimal solution in relation with the optimal solution of the pool-based market. Next, a  decentralized P2P negotiation mechanism and decentralized learning  strategies for prosumers are proposed in Section \ref{distr-trading}. Case studies are then introduced in Section \ref{num} to illustrate the  proposed approaches. Lastly, Section \ref{sum} concludes the paper and provides directions for future research.

%%%%%%%%%%%%%%%%%%%%%%%%%%%%%%%%%%%%%%%%%%%%%%%%%%%%%%%%%%%%%%%%%%%%%%%%%%%%%%%%%%%%%%%%%%%%%%%%%%%%%%%%%%%%%%%%%%%%
\section{P2P Electricity Trading Problem}
\label{sysmod}

Consider the P2P energy trading during the time interval $[1,\bT]$ for a power system consisting of $n$ prosumers, where each prosumer is regarded as a peer or agent who not only consumes power but also can produce power with some sort of power generation or storage. It is assumed that prosumers behave non-strategically, i.e. they do not try to learn the other prosumers' behaviors through the trading process. 

Denote $P_{ij}(t)$ the power to be traded at time step $t$ between the $i$-th and $j$-th prosumers, where $P_{ij}(t)>0$ means prosumer $i$ buys electricity from prosumer $j$, and vice versa, $P_{ij}(t)<0$ means prosumer $i$ sells electricity to prosumer $j$. To simplify the trading of prosumers, it is assumed that at each time step a prosumer only buys or sells power, but not to do both.  In other words, at each time step a prosumer holds only one role, an energy buyer or an energy seller. 

%=====================================
\subsection{Inter-peer Communication Structure}

Denote $\tG(t)$ the inter-prosumer communication graph at time step $t$. Due to the bilateral trading between prosumers, $\tG(t)$ is undirected. Moreover, $\tG(t)$ is a bipartite graph whose node set composes of two disjoint subsets associated to selling and buying prosumers, instead of a fully connected graph (c.f. \cite{Sorin19,Baroche19conf}). Each node only communicates with other nodes in another subset, and do not communicate with nodes inside the same subset. %Thus, $\tG$ is not a fully connected graph (c.f. \cite{Sorin19,Baroche19conf}).  

For each prosumer $i$, denote $\tN_i(t)$ its neighboring set, i.e. the set of other communicated prosumers for energy trading at time step $t$.
Let $E(t)$ denote the incidence matrix of $\tG(t)$ associated with an arbitrary edge orientation. Next, let $a_{ij}$ be elements of the adjacency matrix $\tA(t)$, i.e. $a_{ij}(t)=1$ if prosumers $i$ and $j$ are connected at time step $t$, and  $a_{ij}(t)=0$ otherwise. 
The degree matrix $\tD(t)$ is defined by $\tD(t)=\di\{d_{i}(t)\}_{i=1,\ldots,n}$, where $d_{i}(t) \triangleq \sum_{j \in \tN_{i}(t)}{a_{ij}(t)}$. 
Then the Laplacian matrix $\mathcal{L}(t)$ associated to $\mathcal{G}(t)$ is defined by $\tL(t)=\tD(t)-\tA(t)$.  

%=====================================
\subsection{Objective Function}

Let $n_{i}(t) \triangleq |\tN_i(t)|$, $P_i(t) \in \bR^{n_{i}(t)}$ be the vector of all $P_{ij}(t)$ with $j \in \tN_i(t)$, $P_{i,tr}(t)$ be its total traded power. Then $P_{i,tr}(t)=\mathbf{1}_{n_{i}(t)}^TP_i(t)$.  
Next, denote $C_i(P_{i}(t))$ the total cost of prosumer $i$ for trading in the P2P market, which composes of three components assumed to have the following forms.  
 %of peer $i$, 
\begin{subequations}
\begin{align}
\label{cost-1}
C_{i,1}(P_{i}(t)) &= a_i(t) P_{i,tr}^2(t) + \tilde{b}_i(t) P_{i,tr}(t) \\ %+ c_i(t) 
%\end{equation}
%\begin{equation}
\label{cost-2}
C_{i,2}(P_{i}(t)) &= \sum_{j \in \tN_i} d_{ij} P_{ij}(t) \\
%\end{equation}
%\begin{equation}
\label{cost-3}
C_{i,3}(P_{i}(t)) &= \beta P_{i,tr}(t)
\end{align}
\end{subequations} 
The first component \eqref{cost-1} is an utility function whose parameters $a_i(t)>0$ and $\tilde{b}_i(t)$ are available only for prosumer $i$, which are presented here as time-dependent parameters to reflect the time-varying and complex behaviors of prosumers. 
%However, one may assume that $a_i, \tilde{b}_i, c_i$ are time-invariant for simplicity in analysis. 
The second element \eqref{cost-2} is a bilateral trading cost associated with the traded powers with other prosumers, where $d_{ij}$ is the bilateral trade weight (also called trading coefficient in \cite{Sorin19}) enforced on the trading between prosumer $i$ and prosumer $j$ for the purposes of product differentiation and consumer involvement \cite{Sorin19}. For example, renewable and clean power would be preferred to fossil-based power for reducing emissions of pollutants, hence $d_{ij}$ associated with the former is smaller than that corresponding to the latter. 
%For more details, readers are referred to \cite{Sorin19}. 
The last component \eqref{cost-3} is the implementation cost for the traded powers to be physically executed through the power network, %which is proportional to the amount of traded power, as follows. 
where $\beta>0$ is a fixed rate. %Here, for simplicity in the results presentation, only single-criterion is used. 

Therefore, summing up \eqref{cost-1}--\eqref{cost-3} gives us the following total cost of each prosumer in the P2P market, 
\begin{equation}
\label{cost}
C_{i}(P_{i}(t)) = a_i(t) P_{i,tr}^2(t) + b_i(t) P_{i,tr}(t) + \sum_{j \in \tN_i} d_{ij} P_{ij}(t)
\end{equation} 
where $b_i(t) \triangleq \tilde{b}_i(t) + \beta$. The assumed formula of $C_{i}(P_{i}(t))$ above guarantees that it is convex.

%=====================================
\subsection{System Constraints}

The first constraint is on the bilateral trading power, i.e.,
\begin{equation}
\label{e-1}
P_{ij}(t) + P_{ji}(t) = 0  ~\forall \; j \in \tN_i(t), \ t=1,\ldots,\bT 
\end{equation}
The next constraint is on the limits of power can be traded, %as follows.  
\begin{equation}
\label{e-2}
P_{i,tr}^{\min} \leq P_{i,tr}(t) \leq P_{i,tr}^{\max} ~\forall \; t=1,\ldots,\bT 
\end{equation}
Note that power flow constraints are not considered here %similarly to economic dispatch problems, 
for the sake of simplifying the analytical analysis, and it is assumed that the cost \eqref{cost-3} is paid to the power network operator for dealing with such flow constraints. 

%=====================================
\subsection{Overall Optimization Problem}

The optimal clearing strategy for P2P energy trading is formulated as an optimization problem below. 
\begin{subequations}
\label{p2p}
\begin{align}
\label{cost-p2p}
\min \; & \sum_{t=1}^{\bT} \sum_{i=1}^{n} C_i(P_{i}(t))    \\
\label{balance-p2p}
\textrm{s.t.} ~ & P_{ij}(t) + P_{ji}(t) = 0 ~\forall \; j \in \tN_i(t)  \\
\label{bounded-power-p2p}
~ & P_{i,tr}^{\min} \leq P_{i,tr}(t) = \sum_{j \in \tN_i(t)} P_{ij}(t) \leq P_{i,tr}^{\max} \\
~ & P_{ij}(t) \leq (\geq) \ 0 ~\textrm{if peer $i$ is a seller (buyer)} ~\forall \; j \in \tN_i(t) %\\
%~ & P_{ij}(t) \geq 0 ~\textrm{if peer $i$ is a buyer} ~\forall \; j \in \tN_i(t)
\end{align}
\end{subequations} 
Since the cost functions $C_i(P_{i}(t))$ are convex and the constraints are linear, the mathematical programming \eqref{p2p} is convex. It can be seen that \eqref{p2p} is decomposable with respect to time index, moreover the role of each prosumer can change between a buyer and a seller from time to time, thus hereafter we will solve \eqref{p2p} at each time step and omit the time index.  

In the pool-based market, the optimal energy management problem at each time step $t$ is a social welfare maximization problem, as follows.
\begin{subequations}
\label{swm}
\begin{align}
\label{cost-swm}
\min \; & \sum_{i=1}^{n} C_i(P_{i,tr})    \\
\label{balance-swm}
\textrm{s.t.} ~ & \sum_{i=1}^{n} P_{i,tr} = 0   \\
\label{bounded-power-swm}
~ & P_{i,tr}^{\min} \leq P_{i,tr} \leq P_{i,tr}^{\max} \\
~ & P_{i,tr} \leq (\geq) \ 0 ~\textrm{if peer $i$ is a seller (buyer)} ~\forall \; j \in \tN_i(t) %\\
%~ & P_{i,tr} \geq 0 ~\textrm{if peer $i$ is a buyer} 
\end{align}
\end{subequations} 
Solutions of the P2P market problem \eqref{p2p} and the pool-based market problem \eqref{swm} will be compared to show the differences between these two markets.
\begin{rem}
Time binding constraints, for example 
\begin{equation}
	\label{rate-constr}
	P_{i,tr}^{r,\min}(t) \leq P_{i,tr}(t)-P_{i,tr}(t-1) \leq P_{i,tr}^{r,\max}(t)
\end{equation}
where $P_{i,tr}^{r,\min}(t)$ and $P_{i,tr}^{r,\max}(t)$ are obtained from the predicted generation and consumption of prosumer $i$ at time step $t$, can be included in the P2P optimization problem \eqref{p2p}. Then solving \eqref{p2p} for all time steps at once is suitable for day-ahead or longer scheduling problems. However, this research focuses on the P2P energy trading one-time-ahead, and therefore solve \eqref{p2p} consecutively at each time step. As such, at time step $t$, $P_{i,tr}(t-1)$ is known, hence \eqref{rate-constr} can be rewritten as
$$P_{i,tr}^{r,\min}(t)+P_{i,tr}(t-1) \leq P_{i,tr}(t) \leq P_{i,tr}^{r,\max}(t)+P_{i,tr}(t-1)$$
which is similar to \eqref{bounded-power-p2p}, and therefore can be combined into a unique constraint.  Note that solving P2P optimization problem \eqref{p2p} in real-time would be impractical in real-world systems due to the large scale of the system and possible latencies on inter-peer communications. Thus, the interval between two consecutive time steps is usually an hour or a half hour in real-world systems.
\end{rem}

%=====================================
\subsection{Characterization of Optimal Solution}

Define the following Lagrangian associated to \eqref{p2p},
\begin{equation}
\label{Lagrangian}
L(P_{ij},\lambda_{ij}) = \sum_{i=1}^{n} C_i(P_{i}) - \sum_{i=1}^{n} \sum_{j \in \tN_i} \lambda_{ij}(P_{ij} + P_{ji})  
\end{equation}
where $\lambda_{ij}$ are the Lagrange multipliers associated to the power trading equations \eqref{balance-p2p}, which are regarded as the market clearing prices for energy transactions between pairs of prosumers. 
To essentially compare the solution of the P2P market \eqref{p2p} with the pool-based  market \eqref{swm}, only the equality constraints are considered in the following, since the inequality constraints in those problems are the same. The KKT conditions read as,
\begin{subequations}
\label{kkt}
\begin{align}
\label{kkt-1}
\left. \frac{\partial C_i}{\partial P_{ij}} \right|_{P_{ij}^{\ast}}   &= \lambda_{ij}^{\ast} ~\forall \, j \in \tN_i \\
\label{kkt-2}
P_{ij}^{\ast} + P_{ji}^{\ast}  &= 0 ~\forall \, j \in \tN_i 
\end{align}
\end{subequations}
Condition \eqref{kkt-1} leads to 
\begin{equation}
\label{opt-Lagr-mtl}
2a_i P_{i,tr}^{\ast} + b_i + d_{ij} = \lambda_{ij}^{\ast} ~\forall \, j \in \tN_i
\end{equation}
which shows that the trade weights $d_{ij}$ are parts of the prices $\lambda_{ij}$. 
Denote
\begin{align*}
P_{tr}^{\ast} \triangleq [ 2a_1P_{1,tr}^{\ast},\ldots,2a_nP_{n,tr}^{\ast} ]^T, ~ 
\alpha \triangleq \left[\frac{1}{2a_1},\ldots,\frac{1}{2a_n}\right]^T
\end{align*}

\begin{thm}
\label{opt-char}
The following statements hold. 
\begin{itemize}
	\item[(i)] If no trade weights are imposed, i.e. $d_{ij}=0 ~\forall \, i,j=1,\ldots,n$, and the communication graph between successfully traded peers is connected, then all transaction prices are the same and equal to
\begin{equation}
\label{eq-price}
\lambda^{\ast} = \frac{\sum b_j/(2a_j)}{\sum 1/(2a_j)}
\end{equation}
The optimal total trading power for each peer is
\begin{equation}
\label{eq-power}
P_{i,tr}^{\ast} = \frac{\sum b_j/(2a_j)}{2a_i\sum 1/(2a_j)} - \frac{b_i}{2a_i}
\end{equation}
The sub-indexes in \eqref{eq-price} and \eqref{eq-power} are taken for successfully traded peers.  
Moreover, these optimal price and total traded powers are the same with the optimal solutions of the social welfare maximization problem \eqref{swm}. 
\item[(ii)] If $d_{ij}=0 ~\forall \, i,j=1,\ldots,n$, and the communication graph between successfully traded peers is unconnected, then the considering P2P market is clustered into smaller P2P markets, each of them has a different energy price. The optimal price and traded powers for each smaller P2P market are calculated similarly to \eqref{eq-price} and \eqref{eq-power}.
\item[(iii)] If $d_{ij} \neq 0 $, then energy prices for successful transactions are different from each other. The optimal total traded powers are computed from
\begin{equation}
\label{eq-power-1}
[E,\alpha]^T P_{tr}^{\ast} = [ vec(b_j + d_{ji} - b_i - d_{ij})^T,0]^T
\end{equation}
\end{itemize}
\end{thm}

{\it Proof.} See Appendix.

It is worth emphasizing that the above classical Lagrangian method does not give us a way to compute the individual traded power $P_{ij}$ in each transaction, instead it only provides the total traded power $P_{i,tr}$. Further, as seen in Theorem \ref{opt-char}, the optimal energy prices and trading powers are dependent only on successfully traded prosumers, which, in realistic situations, could be a subset of all participated prosumers. Thus, in the next section, a decentralized approach is proposed to analytically derive the power amount $P_{ij}$ and energy price in each successful transaction in P2P energy market. Then decentralized learning methods are proposed to tune the prosumers cost function parameters such that all of them can successfully trade with desired power amounts. 
%Furthermore, this approach is distributed to naturally fit into the practical implementation by each peer and without any intermediate.

%%%%%%%%%%%%%%%%%%%%%%%%%%%%%%%%%%%%%%%%%%%%%%%%%%%%%%%%%%%%%%%%%%%%%%%%%%%%%%%%%%%%%%%%%%%%%%%%%%%%%%%%%%%%%%%%%%%%
\section{P2P Energy Negotiation Mechanism and Prosumer Learning Strategy}
\label{distr-trading}

\subsection{{\color{blue}Decentralized} P2P Negotiation Mechanism}

In the following, a {\color{blue}decentralized} ADMM approach is proposed to solve the mathematical programming \eqref{p2p} {\color{blue}in parallel}. The advantage of this approach is that it allows each prosumer (peer) to solve its own local optimization problem while negotiating with other prosumers to eventually reach the solution of the global optimization problem \eqref{p2p}. Thus, the communication and computation burden at a central entity is avoided, and the privacy of each prosumer can be better guaranteed. 

It should be noted that the classical two-block ADMM method \cite{Boyd:2011} is centralized (could be implemented in a distributed manner), and the updates of variables are in order. On the other hand, the existing multi-block ADMM approach \cite{Deng2017} allows variables to be updated in parallel, but requires fully connected inter-agent communication which is not fitted into the bipartite structure of the considering P2P energy system. Therefore, in this research, a novel decentralized ADMM approach is proposed for the P2P energy market that solves \eqref{p2p-1} in parallel at all prosumers, and local variables at each prosumer are also updated in parallel. 

Denote $m \triangleq \sum_{i=1}^{n} n_{i}$, $P \in \bR^{m}$ the vector of all $P_i,i=1,\ldots,n$, and the sets of equality constraint \eqref{balance-p2p} and inequality constraint \eqref{bounded-power-p2p} as in \eqref{eq-set} and \eqref{ineq-set}, respectively. 
\begin{align}
\label{eq-set}
\Omega_{eq} \triangleq & \left \{ P \in \bR^{m}: P_{ij}(t) + P_{ji}(t) = 0 ~\forall \, j  \in \tN_i \right \} 
\end{align}
\begin{align}
\label{ineq-set}
\Omega_{ineq} \triangleq & \left \{ P \in \bR^{m}: P_{i,tr}^{\min} \leq \mathbf{1}_{n_i}^T P_i \leq P_{i,tr}^{\max}  \right \}
%& \left. ~ \, b_{i,fc} X_{i,1} - X_{i,2} \leq HW_{i}^{\max}-HT_i(t-1),  \right. \notag \\
%& \quad X_{i,3} \geq 0, \notag \\
%& \quad P_{i,fc}^{\min} \leq X_{i,1} \leq P_{i,fc}^{\max}, \notag \\
%& \quad P_{i,fc}(t-1)+\Delta P_{i,fc}^{\min} \leq X_{i,1}, \notag \\
%& \quad X_{i,1} \leq P_{i,fc}(t-1)+\Delta P_{i,fc}^{\max}, \notag \\
%& \left. ~ \, P_i(t)-P_{i,tr}^{\max} \leq X_{i,1} \leq P_i(t)-P_{i,tr}^{\min} \right \}
\end{align}
For those sets, the following indicator functions are defined. 
\begin{align}
\label{I-eq}
I_{eq}(P) \triangleq \left \{ 
\begin{array}{rl}
	0: & P \in \Omega_{eq} \\
	+\infty: & P \notin  \Omega_{eq}
\end{array}
\right.
\end{align}
\begin{align}
\label{I-ineq}
I_{ineq}(P) \triangleq \left \{ 
\begin{array}{rl}
	0: & P \in \Omega_{ineq} \\
	+\infty: & P \notin  \Omega_{ineq}
\end{array}
\right.
\end{align}
Now, by utilizing a new variable $X \in \bR^{m}$, the optimization problem \eqref{p2p} is rewritten such that equality and inequality constraints are separated into different sets corresponding to different variables $P$ and $X$, as follows. 
\begin{subequations}
\label{p2p-1}
\begin{align}
\min \, & \sum_{i=1}^{n}  C_i(P_i) + I_{eq}(P) + I_{ineq}(X) \\
\textrm{s.t.} \; & P - X = 0 \\
\; & P \in \Omega_{eq}, ~ X \in \Omega_{ineq}
\end{align}
\end{subequations}  
Obviously, \eqref{p2p-1} is in the standard form of the ADMM method \cite{Boyd:2011}. 
Define the following augmented Lagrangian, %associated with (\ref{p2p-1}) as follows,
\begin{align*}
	%\label{aug-Lagrangian}
		L_{\rho}(P,X,u) = & \sum_{i=1}^{n} C_i(P_i) + I_{eq}(P) + I_{ineq}(X)   \\ 
		& +\frac{\rho}{2}\|P-X+u\|_{2}^{2} 
\end{align*}
where $\rho>0$ is a scalar penalty parameter and $u \in \mathbb{R}^{m}$ is called the scaled Lagrange (dual) multiplier \cite{Boyd:2011}. 
Next, the variables $P,X,u$ are computed in parallel at each algorithm iteration $k+1$ by solving the following sub-problems,
	\begin{align}
		\label{var-update}
		X^{k+1} &\triangleq \argmin_{X \in \Omega_{ineq}}{L_{\rho}(P^{k},X,u^{k})  + \frac{1}{2}(X-X^{k})^{T}\Psi(X-X^{k})} \notag \\	
		P^{k+1} &\triangleq \argmin_{P \in \Omega_{eq}}{L_{\rho}(P,X^{k},u^{k}) + \frac{1}{2}(P-P^{k})^{T}\Phi(P-P^{k})} \notag \\
		u^{k+1} &\triangleq u^{k} + \kappa \rho(P^{k+1}-X^{k+1})  
	\end{align}		
in which $\Phi,\Psi,\kappa>0$ satisfy
\begin{equation}
\label{prox-cond}
	\Phi \succ \rho(\frac{1}{\mu_1}-1)I, ~\Psi \succ \rho(\frac{1}{\mu_2}-1)I, ~\mu_1+\mu_2 < 2-\kappa
\end{equation}
for some $\mu_1>0,\mu_2>0$. Condition \eqref{prox-cond} was proved to be sufficient for the convergence of the above variables update \cite{Deng2017}. Note that the selection of $\Phi,\Psi,\kappa$ to fulfill \eqref{prox-cond} is not unique. One simple way is to let $\Phi=\phi I$, $\Psi=\psi I$ such that 
\begin{equation}
\label{prox-cond-1}
	\phi > \rho(\frac{1}{\mu_1}-1), ~\psi > \rho(\frac{1}{\mu_2}-1), 
	~\mu_1+\mu_2 < 2-\kappa
\end{equation}  
There are also multiple choices of $\mu_1,\mu_2$ and $\kappa$ to satisfy \eqref{prox-cond-1}. For instance, let $\mu_1=\mu_2=0.5$, then \eqref{prox-cond-1} becomes $\phi > \rho,~ \psi > \rho, ~\kappa < 1$, which will be used in the case studies later.

Stopping criteria for the iterative process above are \cite{Boyd:2011} 
\begin{equation}
	\label{terminate-criteria}
		\|r^{k} \|_{2} \leq \epsilon^{\rm pri}, ~
		\|s^{k} \|_{2} \leq \epsilon^{\rm dual}, 
\end{equation}
where $r^{k} \triangleq P^{k}-X^{k}$ and $s^{k} \triangleq -\rho\left(X^{k}-X^{k-1}\right)$ are primal and dual residuals at iteration $k$; 
$\epsilon^{\rm pri} > 0$ and $\epsilon^{\rm dual} > 0$ are primal and dual feasibility tolerances chosen by \cite{Boyd:2011}  
\begin{equation}
	\label{tol}
	\begin{aligned}
		\epsilon^{\rm pri} &= \sqrt{n+m}\epsilon^{\rm abs}+\epsilon^{\rm rel}\max\{ \|P^{k}\|_{2},\|-X^{k}\|_{2} \}, \\
		\epsilon^{\rm dual} &= \sqrt{n+m}\epsilon^{\rm abs}+\epsilon^{\rm rel}\|\rho u^{k}\|_{2}, \\
	\end{aligned}
\end{equation}
Here, $\epsilon^{\rm abs} > 0$ and $\epsilon^{\rm rel} > 0$ are absolute and relative tolerances, set in this research to be $10^{-4}$ and $10^{-3}$, respectively. The verification of \eqref{terminate-criteria} in a decentralized manner were presented in \cite[Theorem 2]{Nguyen-TSG17}, hence is omitted here for brevity. 

The updates of variables $P$ and $X$ in \eqref{var-update} are dependent only on their values at the previous iteration, hence can be made without any order, unlike other existing ADMM algorithms for P2P energy systems in the literature, e.g. \cite{KZhang20,Baroche19,Baroche19conf,MorstynP2P19a,Khorasany19}, where variables depend on the others at the current iteration and must be updated one after another.

%===========================================
\subsubsection{The Update for Variable $X$}

The update for $X^{k+1}$ in \eqref{var-update} is derived by solving the following optimization problem.
\begin{subequations}
\label{p2p-X}
\begin{align}
\min \, & \frac{\rho}{2}\|P^k-X+u^k\|_{2}^{2} + \frac{\psi}{2}\|X-X^{k}\|_2^2 \\
\textrm{s.t.} \; & P_{i,tr}^{\min} \leq \mathbf{1}_{n_{i}}^T X_i \leq P_{i,tr}^{\max} ~\forall \, i=1,\ldots,n
\end{align}
\end{subequations} 
Note that there is one more constraint on the positiveness or negativeness of each $X_i,i=1,\ldots,n$, depending on whether prosumer $i$ at the next time slot will perform as a buyer or seller. In any case, \eqref{p2p-X} is a quadratic convex problem and is decomposable to each prosumer/peer, i.e. it is fully decentralized, hence it can be easily solved by any off-the-self software embedded in each prosumer/peer, e.g. CVX \cite{CVX}.

%===========================================
\subsubsection{The Update for Variable $P$}

To obtain the update for $P^{k+1}$ in \eqref{var-update}, we need to solve the following mathematical programming.
\begin{subequations}
\label{p2p-P}
\begin{align}
\min \, & \sum_{i=1}^{n} C_i(P_i) + \frac{\rho}{2}\|P-X^k+u^k\|_{2}^{2} + \frac{\phi}{2}\|P-P^{k}\|_2^2 \\
\label{p2p-balance}
\textrm{s.t.} \; & P_{ij} + P_{ji} = 0 ~\forall \, i=1,\ldots,n; \ j \in \tN_i
\end{align}
\end{subequations}
Denote $\lambda_{ij}>0$ the Lagrange multiplier associated with the constraint \eqref{p2p-balance}, and $\lambda_i \in \bR^{n_{i}}$ the vector of all $\lambda_{ij}$ with $j \in \tN_i$. Since \eqref{p2p-P} is a convex optimization problem with quadratic cost function and linear equality, the strong duality holds and KKT %Karush-Kuhn-Tucker 
conditions apply. Therefore, we obtain from \eqref{p2p-P} that
%\small
\begin{align}
\label{price-eq}
\lambda_{ij}^{k+1}  =& \; \frac{\partial}{\partial P_{ij}} \left(\sum_{i=1}^{n} C_i(P_i) + \frac{\rho}{2}\|P-X^k+u^k\|_{2}^{2} \right. \notag \\
& \qquad \qquad \left. \left. + \frac{\phi}{2}\|P-P^{k}\|_2^2 \right) \right|_{P_{ij}=P_{ij}^{k+1}}  \notag \\
 =& \; 2a_iP_{i,tr}^{k+1} + (\rho+\phi)P_{ij}^{k+1} + v_{ij}^{k}
\end{align}
%\normalsize
where $v_{ij}^{k} \triangleq  b_i + d_{ij} + \rho(-X_{ij}^k+u_{ij}^k) - \phi P_{ij}^k$. 
Next, due to the bilateral trading constraint \eqref{p2p-balance}, the Lagrange multipliers and the traded powers must satisfy the following constraints.
\begin{equation}
\label{price-constraint}
\lambda_{ij}^{k+1} = \lambda_{ji}^{k+1}, P_{ij}^{k+1} = -P_{ji}^{k+1} ~\forall \; j \in \tN_i
\end{equation} 
Here, $\lambda_{ij}^{k+1}$ is considered to be the trading price between the $i$-th and $j$-th prosumers. 
Denote
\begin{align*}
\hat{v}_i^{k+1} &\triangleq \sum_{j \in \tN_i} v_{ji}^{k+1}, ~\tilde{v}_i^{k+1} \triangleq \sum_{j \in \tN_i} v_{ij}^{k+1} \\
\hat{v}^{k+1} &\triangleq \begin{bmatrix} \hat{v}_1^{k+1},\cdots,\hat{v}_n^{k+1} \end{bmatrix}^T, ~ \tilde{v}^{k+1} \triangleq \begin{bmatrix} \tilde{v}_1^{k+1},\cdots,\tilde{v}_n^{k+1} \end{bmatrix}^T \\
\Gamma &\triangleq (\rho+\phi)\mathrm{diag}\{\frac{1}{a_i}\}_{i=1,\ldots,n}
\end{align*}

%Because $\Gamma$ is a positive definite, diagonal matrix, matrix $L+\Gamma$ is always invertible. Thus, we obtain
%\begin{equation}
%\label{opt-P}
%P_{tr}^{k+1} = (\tL + \Gamma)^{-1} \left(\hat{v}^{k+1} - \tilde{v}^{k+1}\right)
%\end{equation}
 
\begin{thm}
\label{opt-char-admm}
The iterative update for individual power and energy transaction price are as follows. 
\begin{subequations}
\begin{align}
\label{opt-var-1}
P_{ij}^{k+1} &= \frac{v_{ji}^{k+1}+2a_jP_{j,tr}^{k+1}-v_{ij}^{k+1}-2a_iP_{i,tr}^{k+1}}{2(\rho+\phi)} \\
\label{opt-var-11}
\lambda_{ij}^{k+1} &= \frac{v_{ji}^{k+1}+2a_jP_{j,tr}^{k+1}+v_{ij}^{k+1}+2a_iP_{i,tr}^{k+1}}{2} 
\end{align}
\end{subequations}
where $P_{i,tr}^{k+1}$ is computed from
\begin{equation}
\label{Ptr}
(\tL + \Gamma) vec(2a_iP_{i,tr}^{k+1}) = \hat{v}^{k+1} - \tilde{v}^{k+1}
\end{equation} 
As $k \rightarrow \infty$ and $\rho \rightarrow 0$, %the proposed algorithm converges, and 
the optimal total traded power and optimal energy price converge to that stated in Theorem \ref{opt-char}. 
\end{thm}

{\it Proof.} See Appendix.

\begin{rem}
Equation \eqref{Ptr} can be solved in a decentralized manner as a decentralized least-square problem using several methods, e.g. \cite{YLiu19,JLiu18}. Details are ignored here for brevity.  
\end{rem}

%===========================================
\subsection{Prosumer Cost Function Parameters Tuning by Learning }
\label{learn-succ-trd}

\subsubsection{Learn For Successful Trading}

As discussed before, it is possible that not all participated prosumers successfully trade. Therefore, this section proposes a simple but effective learning strategy for prosumers (peers) to adjust their cost function parameters to obtain successful energy transaction, when their previous attempt was failed. 

One possible reason for negotiation failure is the dissatisfaction of energy price and the constraint $P_{i,tr}^{\ast}>0 \ (P_{i,tr}^{\ast}<0)$ for buying (selling) prosumer $i$, which lead to $P_{i,tr}^{\ast}=0$. 
Hence, the proposed strategy is based on the analytical formula \eqref{eq-power} %and \eqref{eq-power-1} 
of the optimally traded powers to change prosumers parameters. 
Recalling from \eqref{eq-power} that
\begin{align}
\label{eq-power-new}
P_{i,tr}^{\ast} = \frac{\sum b_j/(2a_j)}{2a_i\sum 1/(2a_j)} - \frac{b_i}{2a_i} %\notag \\
= \frac{\sum (b_j-b_i)/(2a_j)}{2a_i\sum 1/(2a_j)} 
\end{align}
Consequently, the unsuccessfully traded prosumers keep their parameters $a_i$ unchanged, while decreasing (increasing) their parameters $b_i$ if they are buying (selling) prosumers. As such, $P_{i,tr}^{\ast}$ will be increased (decreased) for buying (selling) prosumers. This process is repeated, and the increase or decrease of $b_i$ at each repetition can be selected to be a constant, for simplicity. Then  unsuccessfully bought (sold) prosumers will eventually get $P_{i,tr}^{\ast}>0 \ (P_{i,tr}^{\ast}<0)$, i.e. successful transactions.  

This learning procedure can also be physically explained as follows. The quantity $2a_iP_{i,tr}^{\ast}+b_i$, which is the partial derivative of prosumer $i$ cost function $C_i(\cdot)$ with respect to $P_{i,tr}$ at its optimal value, is often regarded as an optimal marginal cost. Therefore, if prosumer $i$ is a buyer who decreases $b_i$ while keeping $a_i$ unchanged, then his marginal cost decreases, i.e. he can get more profit. Similarly, if prosumer $j$ is a seller who can increase $b_j$ and hold $a_j$ unchanged, then his marginal cost increases, i.e. he is willing to lower his profit. As a result, following the proposed learning method, prosumers will gradually match the expected profits of the others, and achieve successful energy transactions.  

\subsubsection{Learn For Increasing Traded Power Amount}
Employing \eqref{eq-power-new}, there are two possible ways to boost the amounts of traded powers: (i) increase $b_i$; (ii) decrease $a_i$, for instance by a magnitude of $\gamma$, with $\gamma>1$. While both the former and the latter method can boost the traded amounts of powers by $\gamma$ times, the former obviously rises the trading price due to the increase of $v_{ij}$ and $P_{ij}$ in \eqref{opt-var-11}. On the other hand, the latter does not necessarily increase the trading price, because $b_i$ and $a_iP_{ij}$ remain the same in \eqref{opt-var-11}. Thus, the latter method is selected as the learning method for increasing the amounts of traded power in P2P energy market. 

\begin{figure}[htpb!]
		\centering
		\includegraphics[scale=0.35]{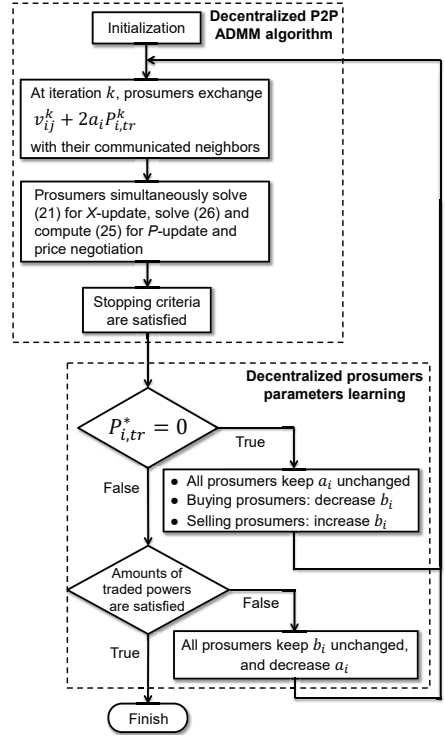}
		%\vspace{-15mm}
		\caption{The proposed distributed P2P ADMM optimization and decentralized prosumers cost function paramters learning approach at a time step.}
		\label{proposed_approach}
	\end{figure}

Finally, the proposed decentralized P2P ADMM algorithm and decentralized cost function parameters learning methods are summarized in Figure \ref{proposed_approach}. 

%%%%%%%%%%%%%%%%%%%%%%%%%%%%%%%%%%%%%%%%%%%%%%%%%%%%%%%%%%%%%%%%%%%%%%%%%%%%%%%%%%%%%%%%%%%%%%%%%%%%%%%%%%%%%%%%%%%%
\section{Case Studies}
\label{num}

%===============================================
\subsection{Synthetic System}
	
This section aims at illustrating the effectiveness of the proposed decentralized optimization algorithm and decentralized learning strategies for P2P energy trading on a synthetic example. A system of 6 prosumers is considered with the parameters given in Table \ref{para}. The parameters $a_i, b_i$ of prosumers are inspired by that of generators reported in the literature (e.g. that in \cite{Nguyen-TSG17} and references therein). P2P energy trading is assumed to occur every hour.
%Later, it is shown how the proposed decentralized learning strategies in Section \ref{learn-succ-trd} can be used to adjust those parameters to obtain desired trades for prosumers.} 

\begin{table}[htpb!]	 
	\caption{Parameters for synthetic system.}
	\begin{center}
		\scalebox{0.97}{
		\begin{tabular}{|c|c|c|c|c|c|c|}
			\hline
			 Prosumer & 1 & 2 & 3 & 4 & 5 & 6  \\
			\hline
			$a_i$ & 0.0031 & 0.0074 & 0.0066 & 0.0063 & 0.0069 & 0.0095 \\
			\hline 							
			$b_i$ & 8.71 & 3.53 & 7.58 & 2.24 & 8.53 & 3.46 \\
			\hline 			
			 $P_{i,tr}^{\min}$ [kW] & -105 & -115 & -125 & 0.01 & 0.01 & 0.01 \\
			\hline 	
			 $P_{i,tr}^{\max}$ [kW] & -0.01 & -0.01 & -0.01 & 100 & 110 & 95 \\
			\hline 								
		\end{tabular}
		}
	\end{center}	
	\label{para}
\end{table}

	\begin{figure}[htpb!]
		\centering
		\includegraphics[scale=0.5]{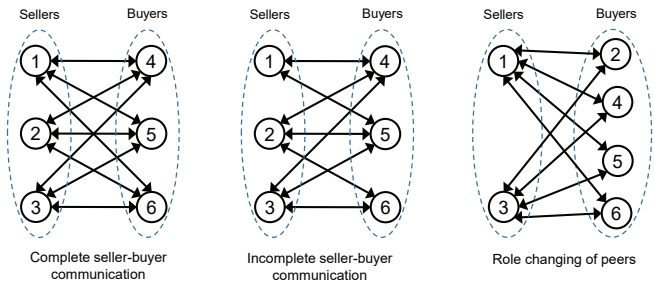}
		%\vspace{-15mm}
		\caption{Time-varying inter-peer communication structure.}
		\label{com-struct}
	\end{figure}	

Six scenarios will be considered. Scenarios 1 examines the solutions of the pool-based market and compares them with the P2P market solutions. Then scenarios 2 and 3 with different inter-peer communication structures will be investigated (see first two graphs in Figure \ref{com-struct}).  Next, in the 4th scenario, results of P2P market when a prosumer changes its role will be presented (see the last graph in Figure \ref{com-struct}). These four scenarios do not consider the bilateral trade weights $d_{ij}$, hence scenario 5 will investigate the effects of such weights. Lastly, scenario 6 demonstrates the proposed decentralized learning strategies in Section \ref{learn-succ-trd} such that all prosumers successfully trade. In all illustrating figures, dash lines represent energy transaction prices, whereas solid lines show traded powers. All simulations are conducted in MATLAB installed on a computer equipped with Intel Core i7-6700K CPU 4GHz and 64GB RAM. Optimization problem \eqref{p2p-X} for variable $X$ update is solved by the software CVX \cite{CVX}. All variables are initialized at zero.

%------------------------------------------
\subsubsection{Scenario 1 (Pool-based market)}

%This section is devoted for the comparison between the results without P2P energy trading and the P2P results in the previous case studies. 

\begin{figure}[htpb!]
		\centering
		\includegraphics[scale=0.3]{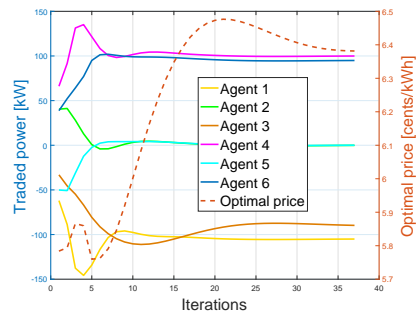}
		%\vspace{-15mm}
		\caption{Power trading in Scenario 1.}
		\label{power_noP2P}
	\end{figure}

Decentralized ADMM approaches for solving the problem \eqref{swm} without P2P energy trading was proposed in our previous works \cite{Nguyen-TSG17,Nguyen19}. The results presented in Figure \ref{power_noP2P} 
are obtained using the algorithm in \cite{Nguyen19} with $\rho=0.02,\phi=0.021,\psi=0.021,\kappa=0.99$.

	%\begin{figure}[htpb!]
		%\centering
		%\includegraphics[scale=0.3]{cost_noP2P}
		%%\vspace{-15mm}
		%\caption{Objective functions of agents in Scenario 1.}
		%\label{cost_noP2P}
	%\end{figure}	
	
%------------------------------------------
\subsubsection{Scenario 2 (P2P market with buyer-seller complete communication)}

 	\begin{figure}[htpb!]
		\centering
		\includegraphics[scale=0.3]{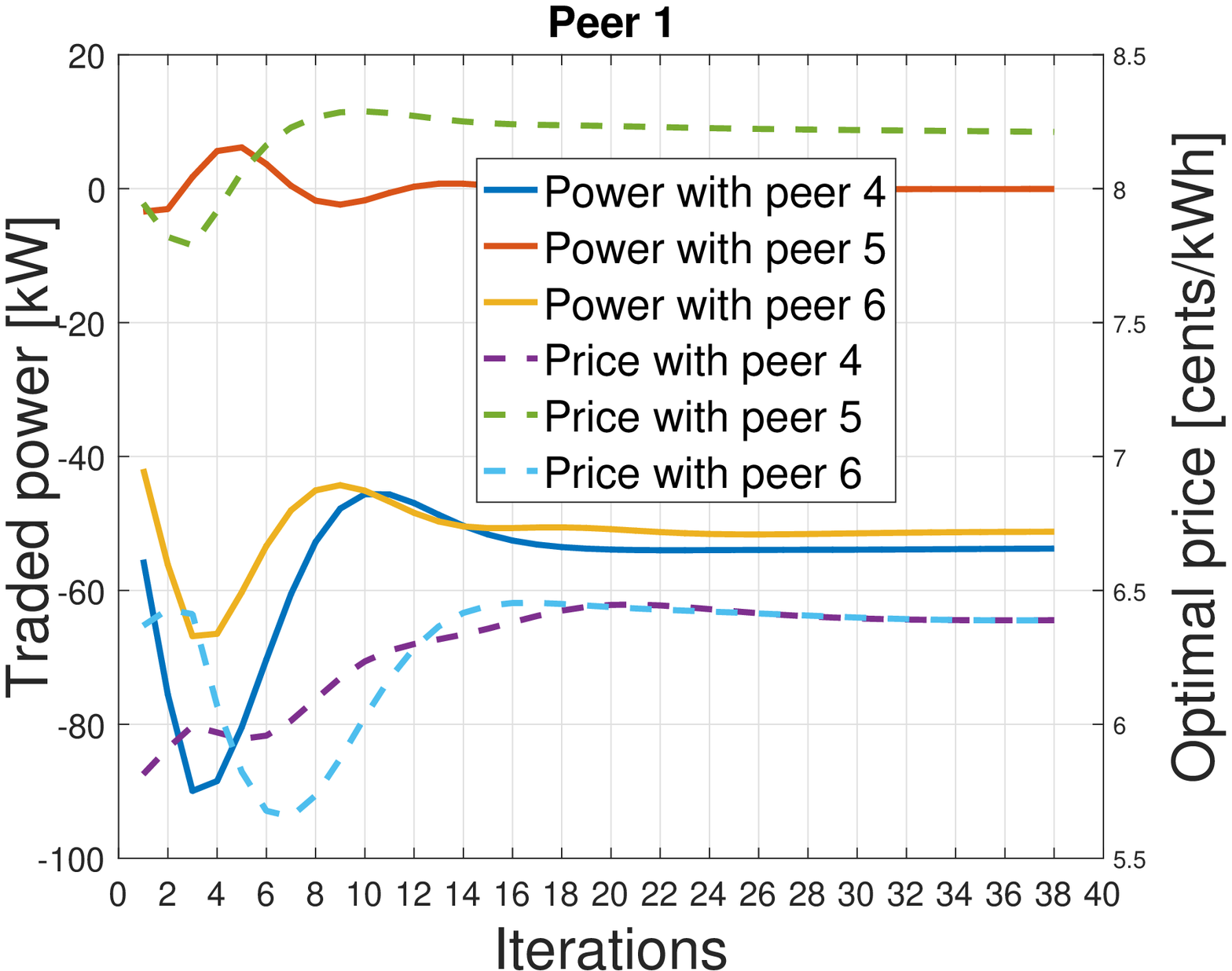}
		\includegraphics[scale=0.3]{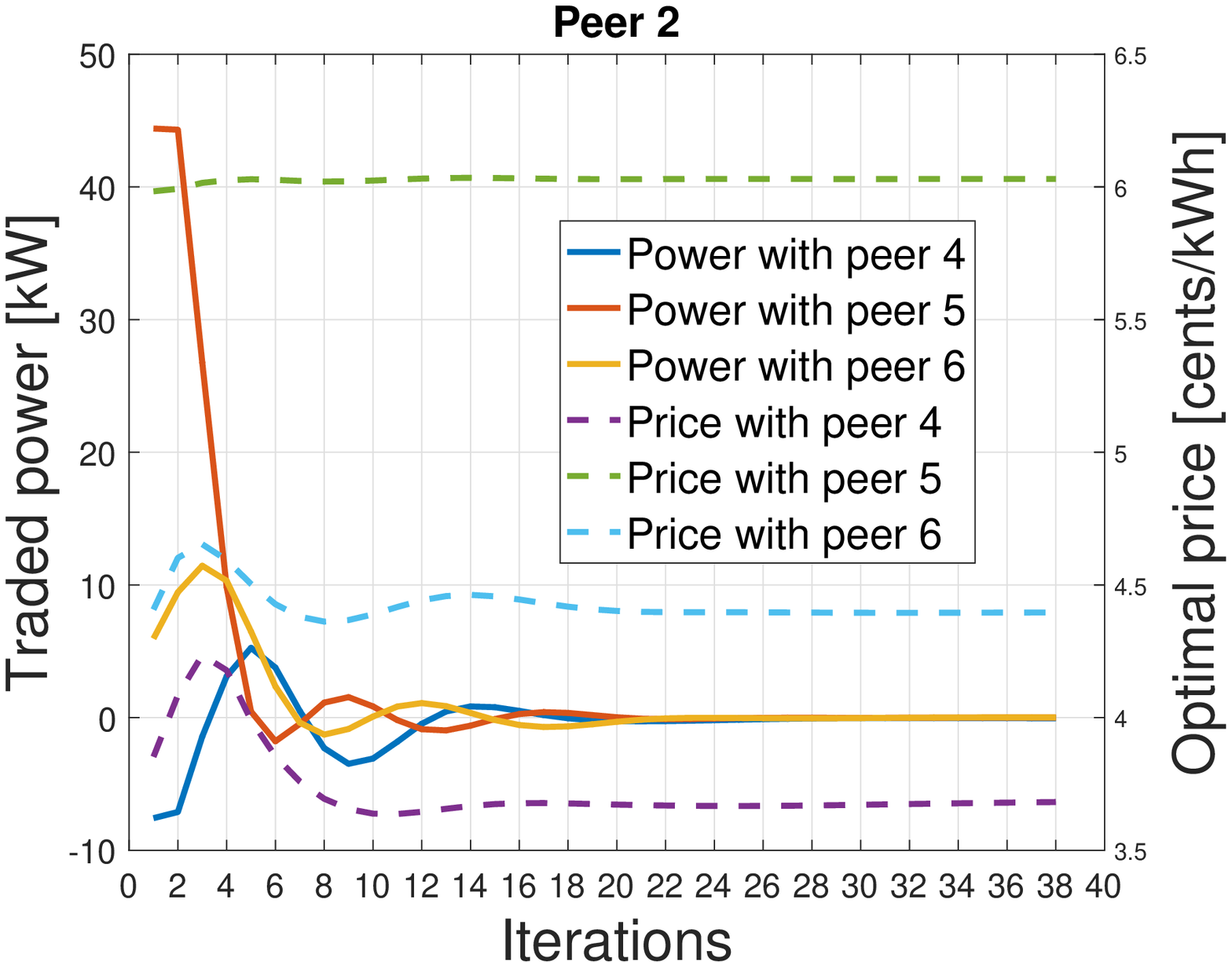}
		\includegraphics[scale=0.3]{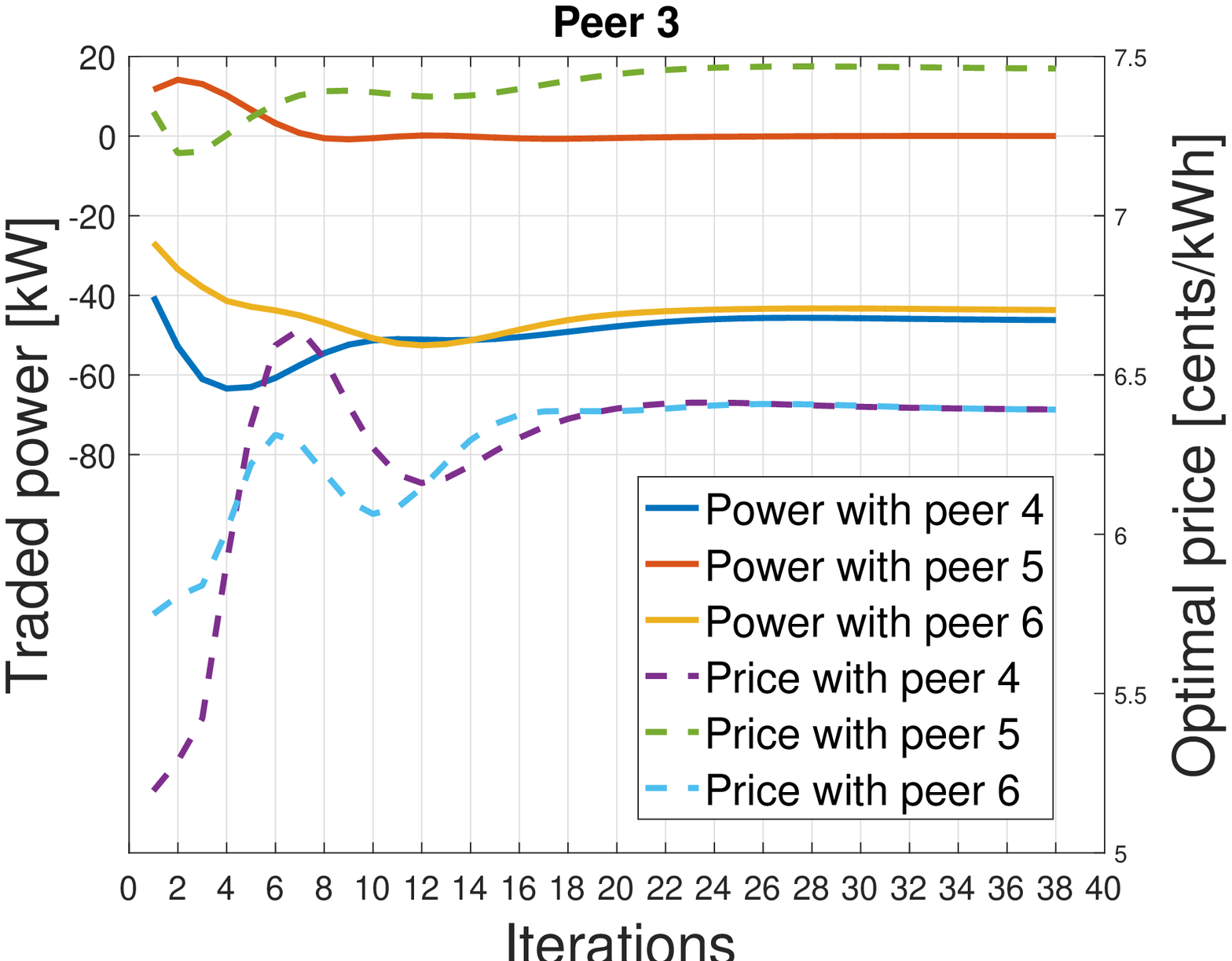}
		%\vspace{-15mm}
		\caption{P2P power trading in Scenario 2.}
		\label{p3_case1}
	\end{figure}	
	%\begin{figure}[htpb!]
		%\centering
		%\includegraphics[scale=0.3]{cost_case1}
		%%\vspace{-15mm}
		%\caption{Objective functions of prosumers in Scenario 2.}
		%\label{cost_case1}
	%\end{figure}	

In this scenario, each buying prosumer is communicated with each selling prosumer, as displayed in the first graph in Figure \ref{com-struct}. Parameters used in the proposed Decentralized ADMM algorithm are the same with scenario 1. Simulation results depicted in Figure \ref{p3_case1} show that peer 2 and peer 5 do not trade, but energy prices of all successful transactions are identical, because the communication graph between successfully traded peers is connected. 
Additionally, the total traded power of each peer and energy price are the same with that obtained in scenario 1. This confirms the first statement of Theorem \ref{opt-char}.

%------------------------------------------------
\subsubsection{Scenario 3 (P2P market with buyer-seller incomplete communication)}

\begin{figure}[htpb!]
		\centering
		\includegraphics[scale=0.3]{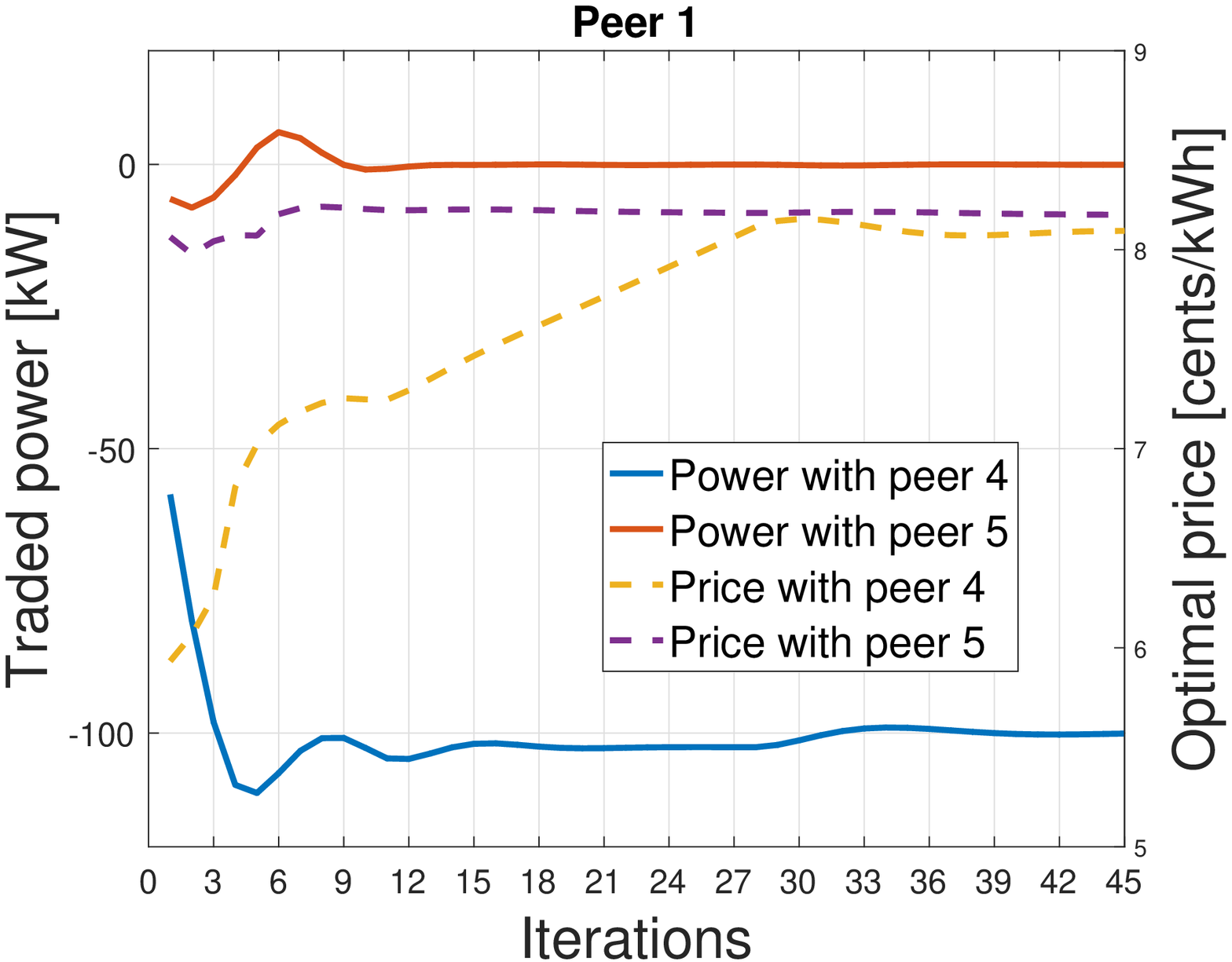}
		\includegraphics[scale=0.3]{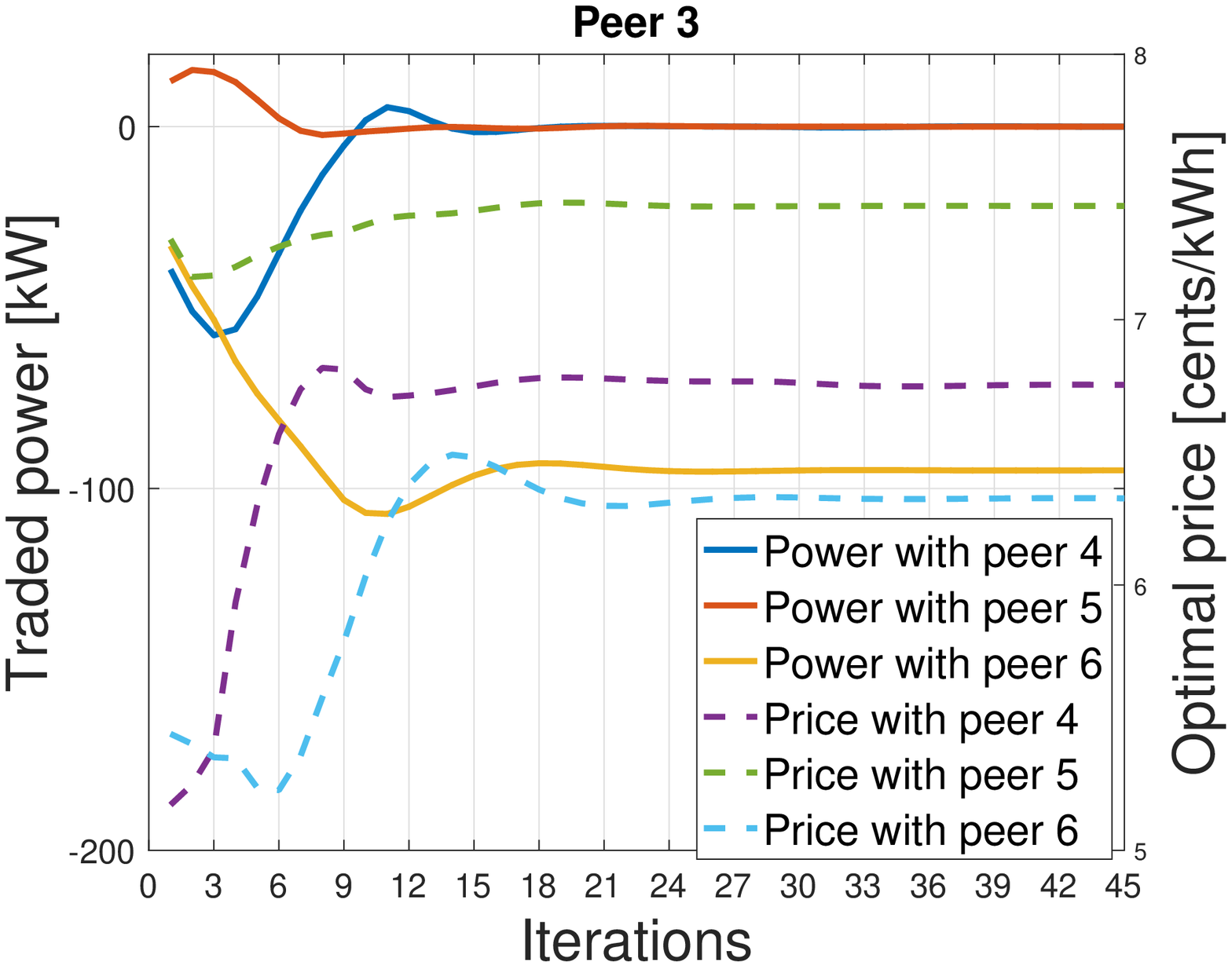}
		%\vspace{-15mm}
		\caption{P2P power trading in Scenario 3.}
		\label{p3_case2}
	\end{figure}

Suppose that the communication link between the 1st and 6th prosumers is removed, as seen in the second graph in Figure \ref{com-struct}. 
ADMM parameters are the same with scenario 2. 
Simulation results for this scenario are then shown in Figure \ref{p3_case2}. Similar to scenario 2, peer 2 and peer 5 also do not successfully trade. 
On the other hand, peer 1 only sells power to peer 4, whilst peer 3 only sells power to peer 6, hence the communication graph between successfully traded peers is unconnected. 
This results in a clustered P2P market consisting of two small P2P markets with different energy prices. This result illustrates the 2nd statement of Theorem \ref{opt-char}.

	%\begin{figure}[htpb!]
		%\centering
		%\includegraphics[scale=0.3]{cost_case2}
		%%\vspace{-15mm}
		%\caption{Objective functions of prosumers in Scenario 3.}
		%\label{cost_case2}
	%\end{figure}	

%---------------------------------------------
\subsubsection{Scenario 4 (P2P market with role changing of peers)}

This section demonstrates the P2P market flexibility and time-varying behaviors of prosumers, where a seller, particularly peer 2, now becomes a buyer. Accordingly, lower and upper bounds for traded powers of peer 2 are reversed with opposite signs. 
ADMM parameters are the same with scenario 2. 

\begin{figure}[htpb!]
		\centering
		\includegraphics[scale=0.3]{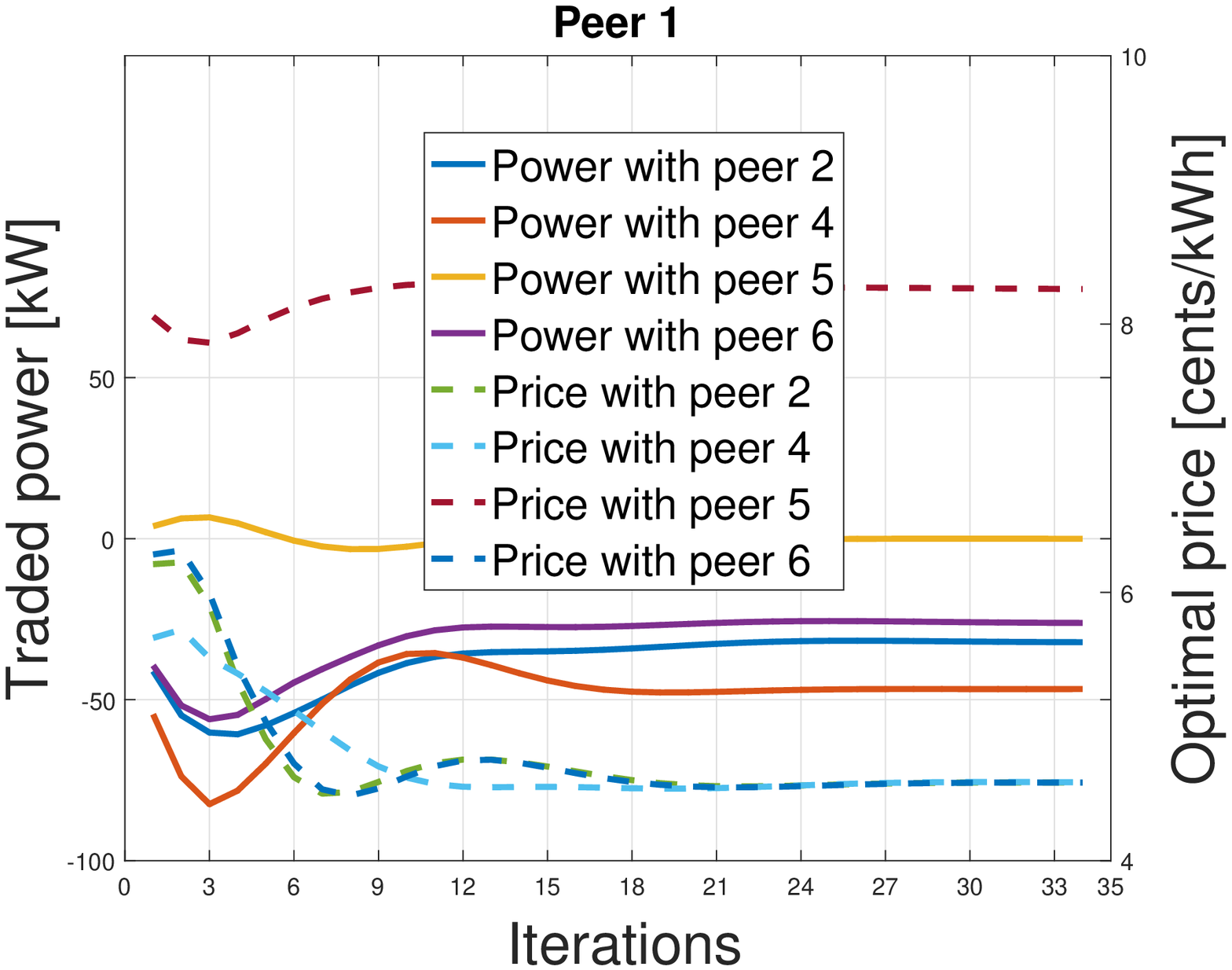}
		\includegraphics[scale=0.3]{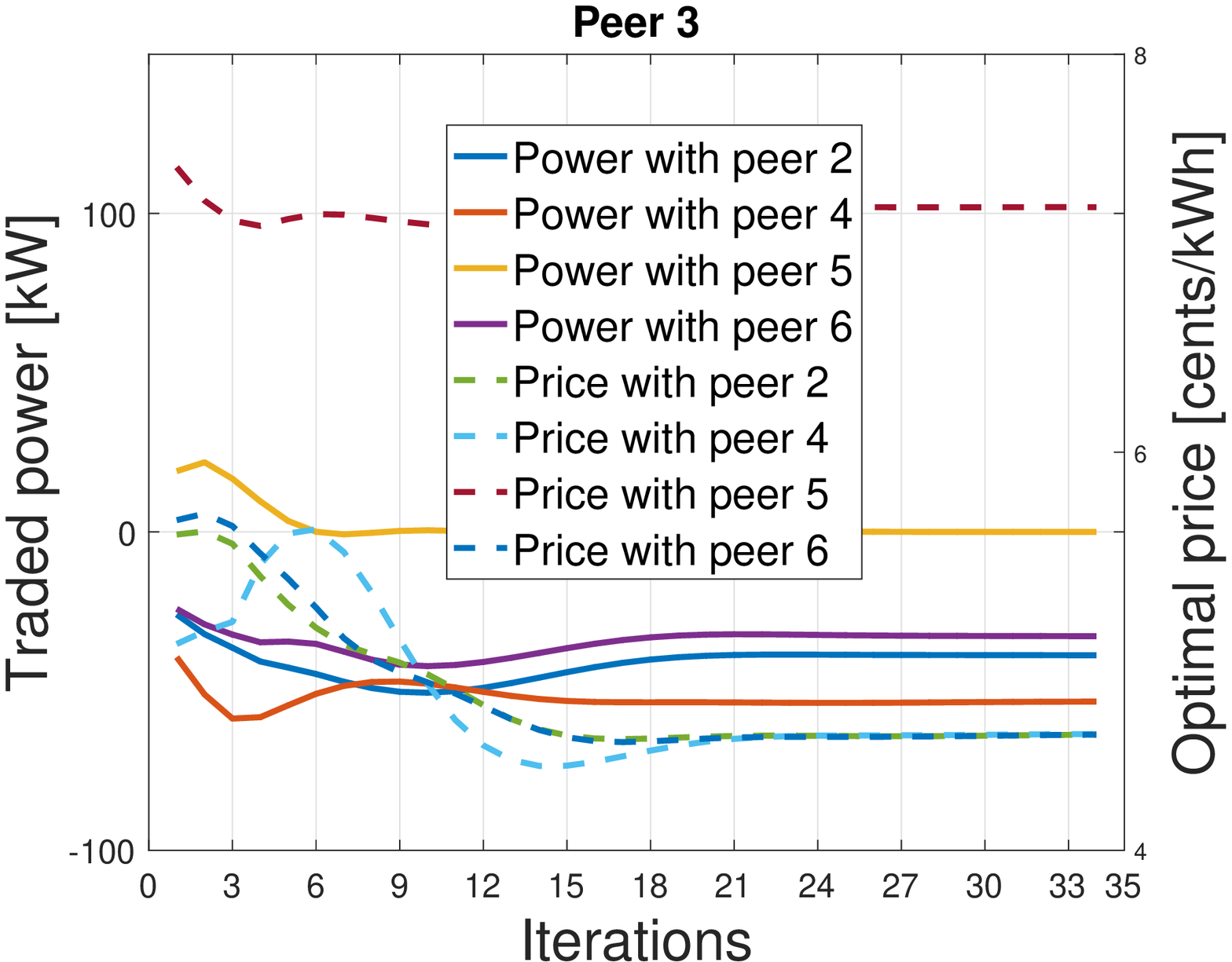}
		%\vspace{-15mm}
		\caption{P2P power trading in Scenario 4.}
		\label{p3_case5}
	\end{figure}

The simulation results are then displayed in Figure \ref{p3_case5}. Peer 5 again does not trade, like in all previous cases, but peer 2 now buys power from both peer 1 and peer 3. Peer 4 and peer 6 also buy power from both peer 1 and peer 3.   
It is noted that the traded price and powers in this scenario are distinct from that in scenarios 2 and 3 because of peer 2 role changing, which illustrate the time-varying behaviors of prosumers in P2P energy market.

	%\begin{figure}[htpb!]
		%\centering
		%\includegraphics[scale=0.3]{cost_case5}
		%%\vspace{-15mm}
		%\caption{Objective functions of prosumers in Scenario 4.}
		%\label{cost_case5}
	%\end{figure}			
	
%---------------------------------------------------
\subsubsection{Scenario 5 (P2P market with effects of bilateral trade weights)}

The communication structure between buying and selling prosumers in this case is the same with that in scenario 2, and single-criterion bilateral trade weights are enforced for each transaction based on the energy sources, i.e. renewable or fossil-based. As such, we assume that peer 1 is fossil-based producer, whilst peer 2 and peer 3 are renewable producers, and $d_{41}=0.51$, $d_{51}=0.51$, $d_{61}=0.72$, $d_{42}=d_{43}=0.1$, $d_{52}=d_{53}=0.12$, $d_{62}=d_{63}=0.04$. These parameters show strong preference of buyers to renewable over fossil-based power, and the strongest supporter is peer 6. 

\begin{figure}[htpb!]
		\centering
		\includegraphics[scale=0.3]{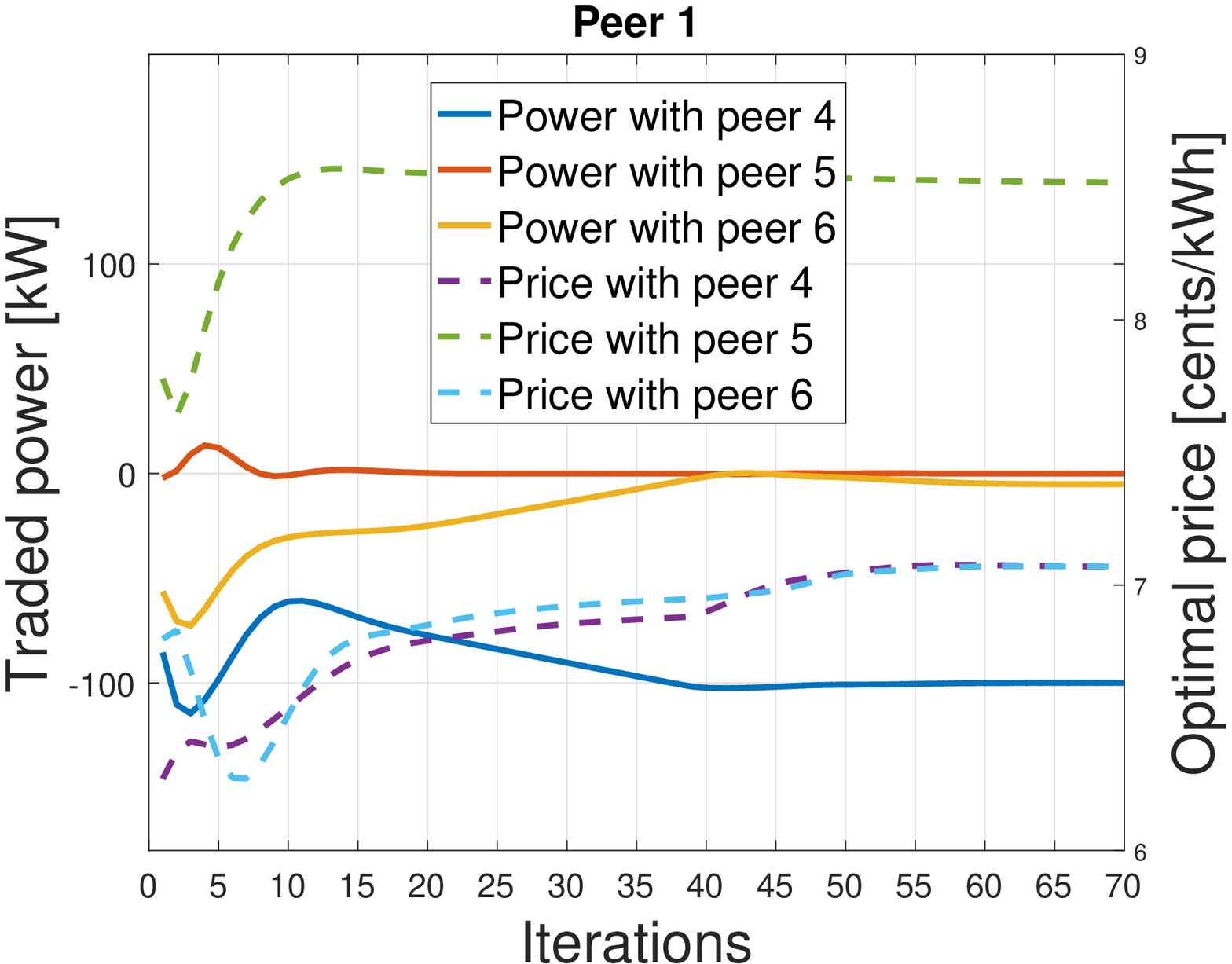}
		\includegraphics[scale=0.3]{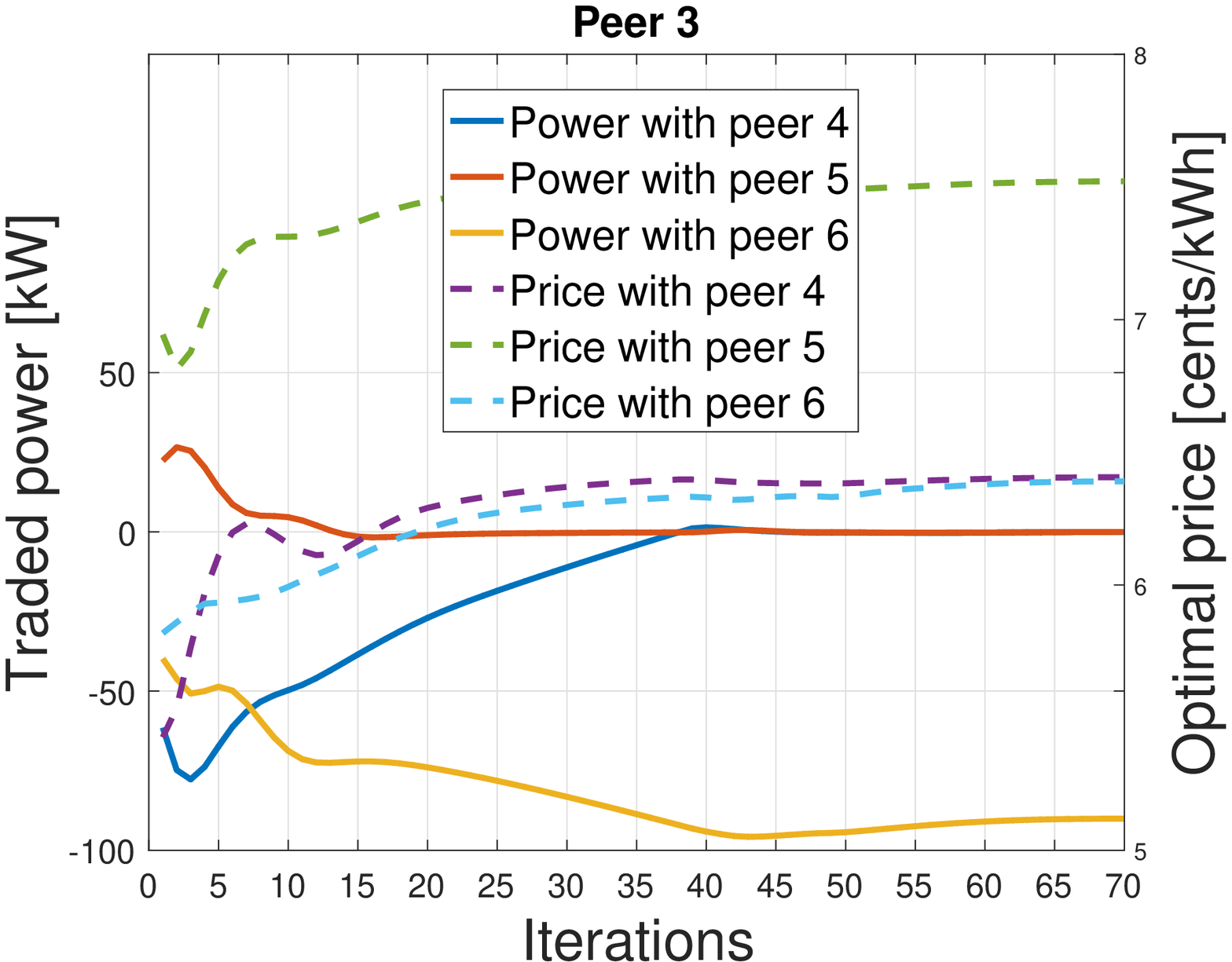}
		%\vspace{-15mm}
		\caption{P2P power trading in Scenario 5.}
		\label{p3_case4}
	\end{figure}

The simulation results in this scenario are then shown in Figure \ref{p3_case4} with $\rho=0.009,\phi=0.0091,\psi=0.0091$ which give heuristically fastest convergence among many different values of $\rho,\phi,\psi$. It is observed that peers 2 and 5 are unsuccessfully traded, like in scenarios 2--4, while results for other peers are substantially changed  because of the trade weights $d_{ij}$, as anticipated in the 3rd statement of Theorem \ref{opt-char}.  

First, the energy transaction prices are no longer the same as in Scenario 2, even the buyer-seller communication structure is complete. 
Second, the trading of each buyer-seller pair are completely different from that in previous cases. Particularly, the power that peer 6 trades with peer 1 becomes very small, 5.1 kW, in comparison with 51.2 kW in Scenario 2. Peer 6 then turns to peer 3 to buy 90.1 kW from it. This is fully explainable due to the strong opposition of peer 6 to fossil-based generation (peer 1) while showing high interest in renewable supply (peer 3). Hence, in realistic P2P energy markets, trade weights can be employed by prosumers to help attain strategic objectives such as emissions reduction, loss decrease, etc., for example by putting small values for renewable and big values for fossil-based sources.

	%\begin{figure}[htpb!]
		%\centering
		%\includegraphics[scale=0.3]{cost_case4}
		%%\vspace{-15mm}
		%\caption{Objective functions of prosumers in Scenario 5.}
		%\label{cost_case4}
	%\end{figure}		

\begin{table}[htpb!]	
	\caption{Convergence time for synthetic system.}
	\begin{center}
		\scalebox{1}{
		\begin{tabular}{|c|c|c|c|c|c|c|}
			\hline
			 Scenario & 1 & 2 & 3 & 4 & 5 & 6  \\
			\hline
			Time & 0.008s & 11s & 10.6s & 7.9s & 17.9s & 10.3s \\
			\hline 							
		\end{tabular}
		}
	\end{center}	
	\label{table-1}
\end{table}

\begin{table*}[htpb!]	
	\caption{Total traded power (in black) and costs (in red) of prosumers in different scenarios.}
	\begin{center}
		\scalebox{1}{
		\begin{tabular}{|c|c|c|c|c|c|c|}
			\hline
			 Pros. & Scen.1 & Scen.2 & Scen.3 & Scen.4 & Scen.5 & Scen.6  \\
			\hline
			\multirow{2}{*}{1} & -105 & -105 & -100 & -105 & -105 & -105 \\
			\cline{2-7} 					
			& \textcolor{red}{-669.9} & \textcolor{red}{-669.9} &  \textcolor{red}{-810} &  \textcolor{red}{-480.9} & \textcolor{red}{-742.35} & \textcolor{red}{-649.95} \\
			\hline
			 \multirow{2}{*}{2} & 0 & 0 & 0 & 70.93 & 0 & -96.13 \\
			\cline{2-7} 					
			& \textcolor{red}{0} & \textcolor{red}{0} & \textcolor{red}{0} & \textcolor{red}{-480.9} & \textcolor{red}{0} & \textcolor{red}{-595.05} \\			
			\hline 
			 \multirow{2}{*}{3} & -90 & -90 & -95 & -124.83 & -90 & -103.87 \\
			\cline{2-7} 
			& \textcolor{red}{-574.2} & \textcolor{red}{-574.2} & \textcolor{red}{-600.97} & \textcolor{red}{-571.72} & \textcolor{red}{-636.3} & \textcolor{red}{-642.96} \\					
			\hline 
			 \multirow{2}{*}{4} & 100 & 100 & 100 & 100 & 100 & 100 \\
			\cline{2-7} 
			& \textcolor{red}{638} & \textcolor{red}{638} & \textcolor{red}{810} & \textcolor{red}{458} & \textcolor{red}{707} & \textcolor{red}{619} \\	
			\hline 
			 \multirow{2}{*}{5} & 0 & 0 & 0 & 0 & 0 & 110 \\
			\cline{2-7}
			& \textcolor{red}{0} & \textcolor{red}{0} & \textcolor{red}{0} & \textcolor{red}{0} & \textcolor{red}{0} & \textcolor{red}{680.9} \\	
			\hline 
			 \multirow{2}{*}{6} & 95 & 95 & 95 & 58.9 & 95 & 95 \\
			\cline{2-7}
			& \textcolor{red}{606.1} & \textcolor{red}{606.1} & \textcolor{red}{600.97} & \textcolor{red}{269.76} & \textcolor{red}{671.65} & \textcolor{red}{588.05} \\	
			\hline 																	
		\end{tabular}
		}
	\end{center}	
	\label{table-2}
\end{table*}

%---------------------------------------------------
\subsubsection{Scenario 6 (Decentralized learning for successful trading)}

This section illustrates the decentralized learning strategy for obtaining successful trading proposed in Section \ref{learn-succ-trd}. The inter-peer communication structure is the same with that in scenario 2. Since peer 2 and peer 5 were failed to trade with other peers in all previously introduced scenarios, the proposed learning strategy is only applied to them. Accordingly, $a_2$ and $a_5$ are kept unchanged, whereas $b_2$ is increased and $b_5$ is decreased. 

\begin{figure}[htpb!]
		\centering
		\includegraphics[scale=0.3]{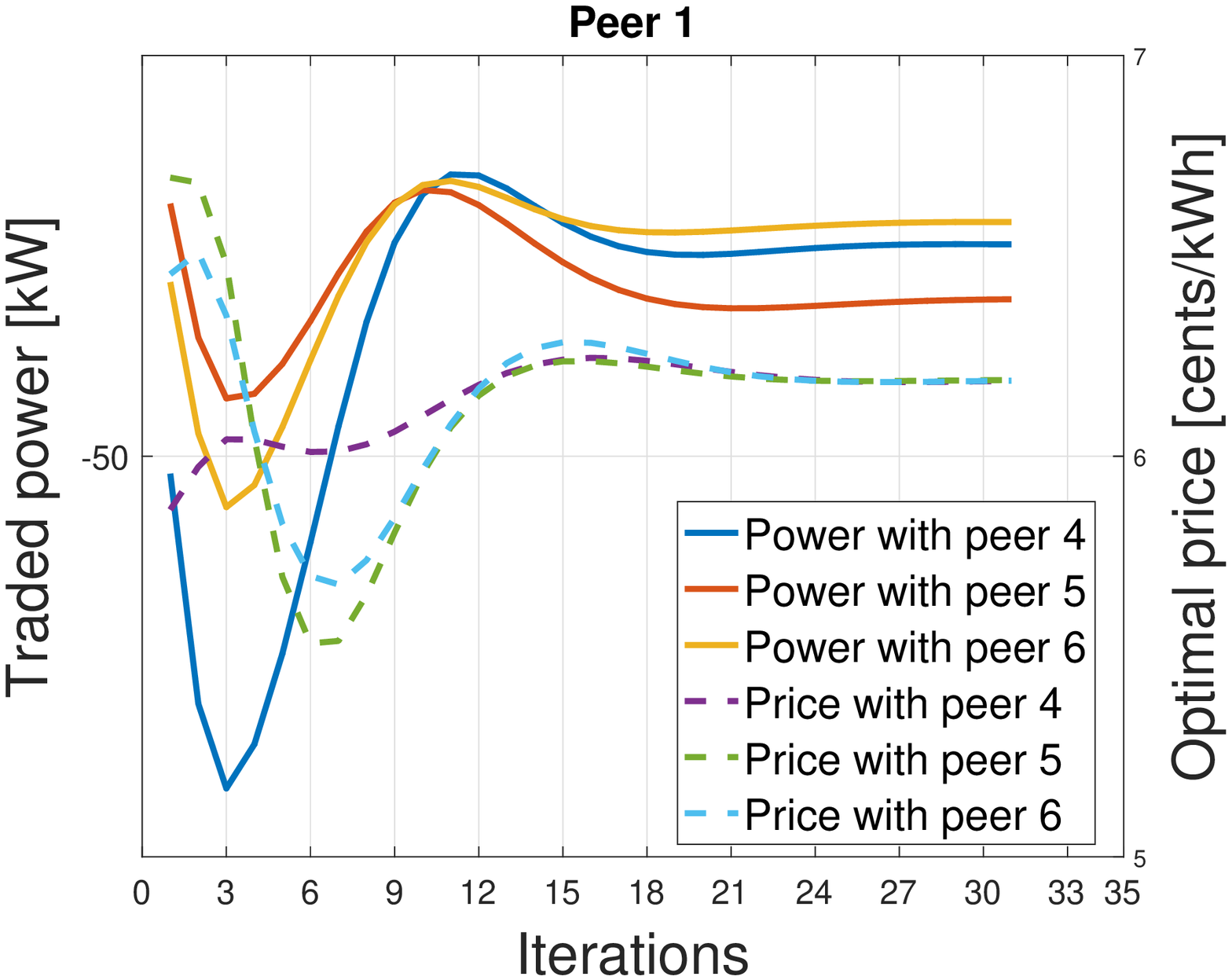}
		\includegraphics[scale=0.3]{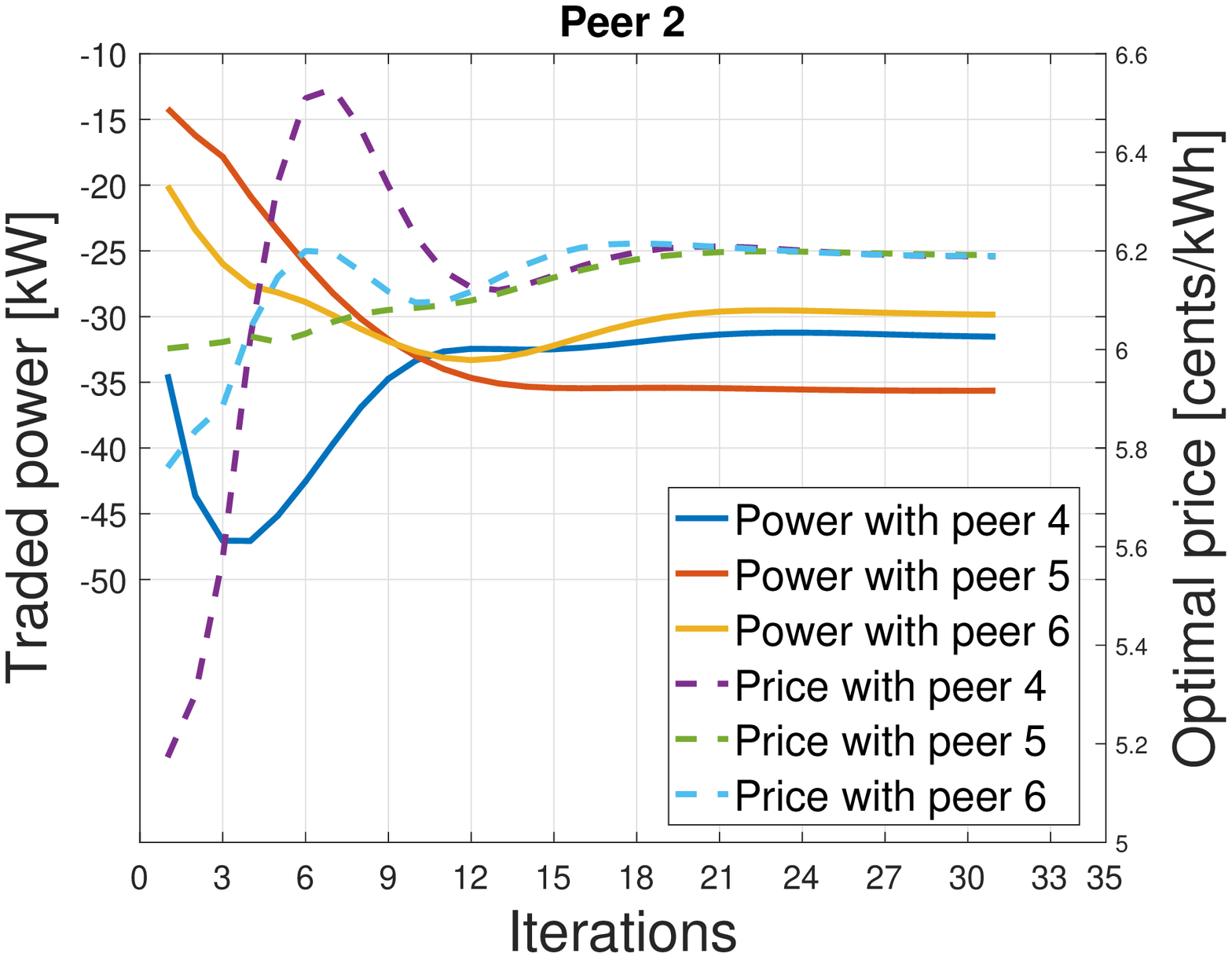}
		\includegraphics[scale=0.3]{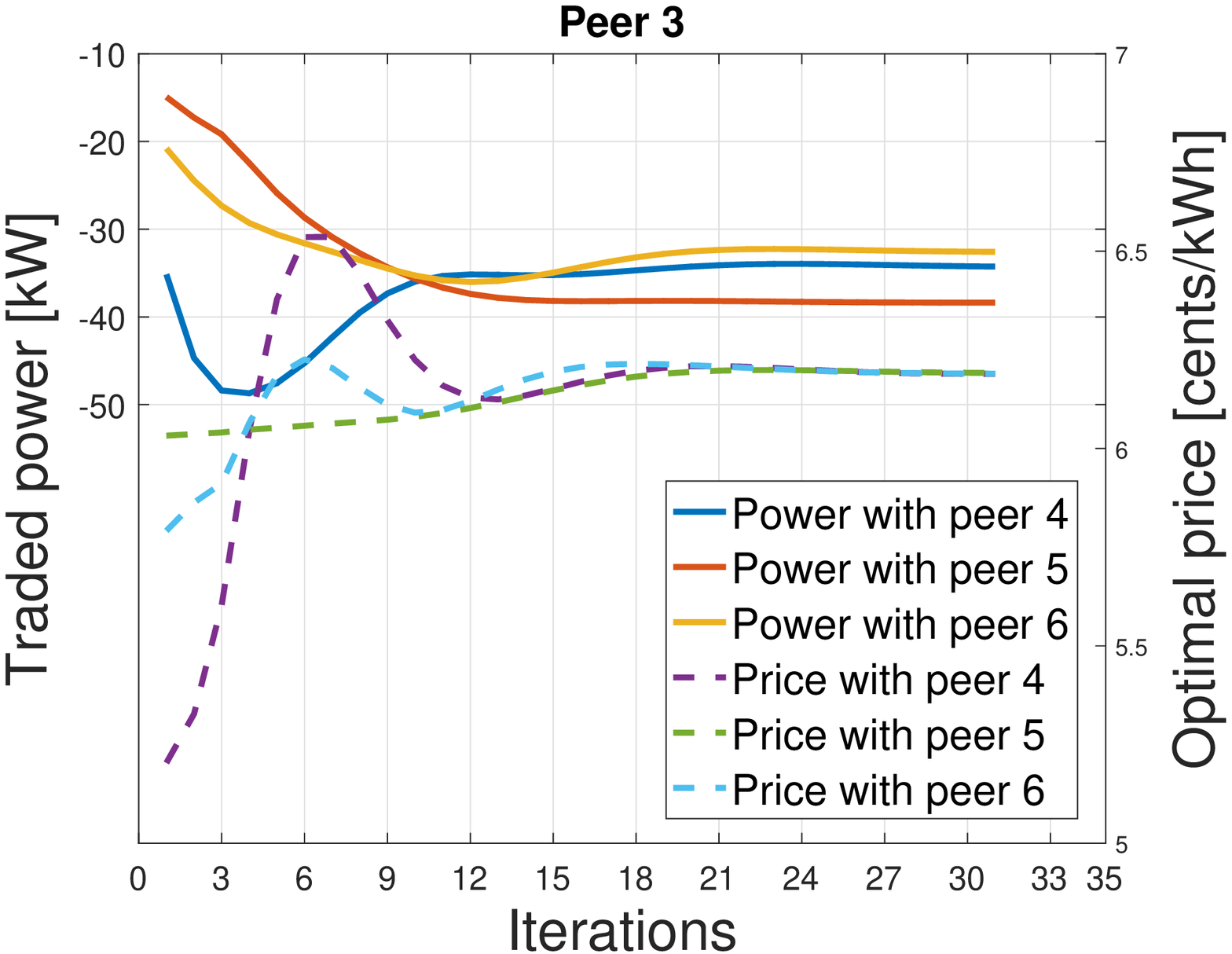}
		%\vspace{-15mm}
		\caption{P2P power trading in Scenario 6.}
		\label{p3_case6}
	\end{figure}

The simulation results exhibited in Figure \ref{p3_case6} are obtained when $b_2=7.53$ and $b_5=4.53$. It can be seen that all peers now successfully trade, which demonstrates the efficiency of the proposed learning strategy. 

	%\begin{figure}[htpb!]
		%\centering
		%\includegraphics[scale=0.3]{cost_case6}
		%%\vspace{-15mm}
		%\caption{Objective functions of prosumers in Scenario 6.}
		%\label{cost_case6}
	%\end{figure}		

Next, convergence time (without inter-peer communication time) of the proposed decentralized ADMM algorithm for P2P trading in all scenarios is shown in Table \ref{table-1}. It can be observed that computational time of scenario 5 with trade weights is longest, while that of scenario 1 with pool-based market is fastest, and that of the others are quite similar. This is logical because in pool-based market, no convex optimization problem is needed to solve in the $X$-update step, unlike that in the P2P market, and trade weights make bilateral trading asymmetric leading to more time for convergence.

Finally, total traded powers and costs for prosumers across all scenarios are provided in Table \ref{table-2}. Those for scenarios 1 and 2 are the same, as seen before. Total traded costs in scenarios 3 and 4 are significantly different from other cases, as the system is clustered (scenario 3), or a prosumer switches between selling and buying roles (scenario 4). Total traded costs in scenario 6 reveal that prosumers 4 and 6 need to pay less, whereas prosumer 1 suffers a bit 
loss, and prosumer 3 gains considerably more, compared to scenarios 1 and 2. Hence, the decentralized learning for successful trading of prosumers 2 and 5 not only helps themselves but also is beneficial for many other prosumers, though not all. On the other hand, energy price is increased in scenario 5 though trade weights help change the traded amounts of powers, leading to higher payments for prosumers 4 and 6, and simultaneously bring more profit to prosumer 1 and 3. Thus, how to choose the best trade weights to enforce on bilateral trading between prosumers needs further investigation.

%===============================================
\subsection{Modified IEEE European Low Voltage Test Feeder}
\label{eu-test}

	\begin{figure}[htpb!]
		\centering
		\includegraphics[scale=0.3]{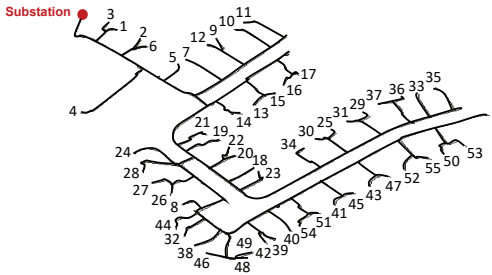}
		%\vspace{-15mm}
		\caption{Diagram of the IEEE European Low Voltage Test Feeder.}
		\label{eu_55nodes}
	\end{figure}

\begin{figure}[htpb!]
		\centering
		\includegraphics[scale=0.3]{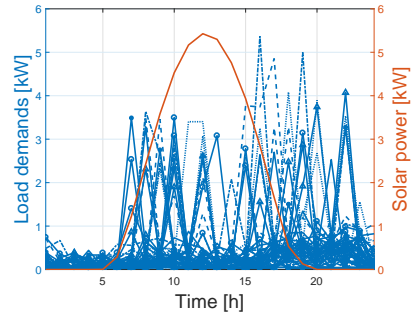}
		%\vspace{-15mm}
		\caption{Load demands and assumed solar generation in the IEEE European Low Voltage Test Feeder.}
		\label{eu_1h_loads_solar}
	\end{figure}

In this section, validation of the proposed approach is carried out for the IEEE European Low Voltage Test Feeder having 55 nodes \cite{eu_55} shown in Figure \ref{eu_55nodes}. It is assumed that 25 nodes have 5.5kW rooftop solar systems, while the remaining nodes have 3kWh battery systems. P2P energy trading is allowed in this system each one-hour interval. One-hour load profiles of 55 nodes are taken from \cite{eu_55}, and average daily global solar irradiance data in July in Spain are obtained from \cite{eu_solar_data}. Then power output of 5.5kW solar systems are computed using the formula in \cite{Riffonneau11} (see Eq. (6) in \cite{Riffonneau11}), where temperature effect (see Eq. (7) in \cite{Riffonneau11}) is ignored because it is very small. Load demands of all 55 nodes and solar power output are depicted in Figure \ref{eu_1h_loads_solar}. 

	\begin{figure}[htpb!]
		\centering
		\includegraphics[scale=0.3]{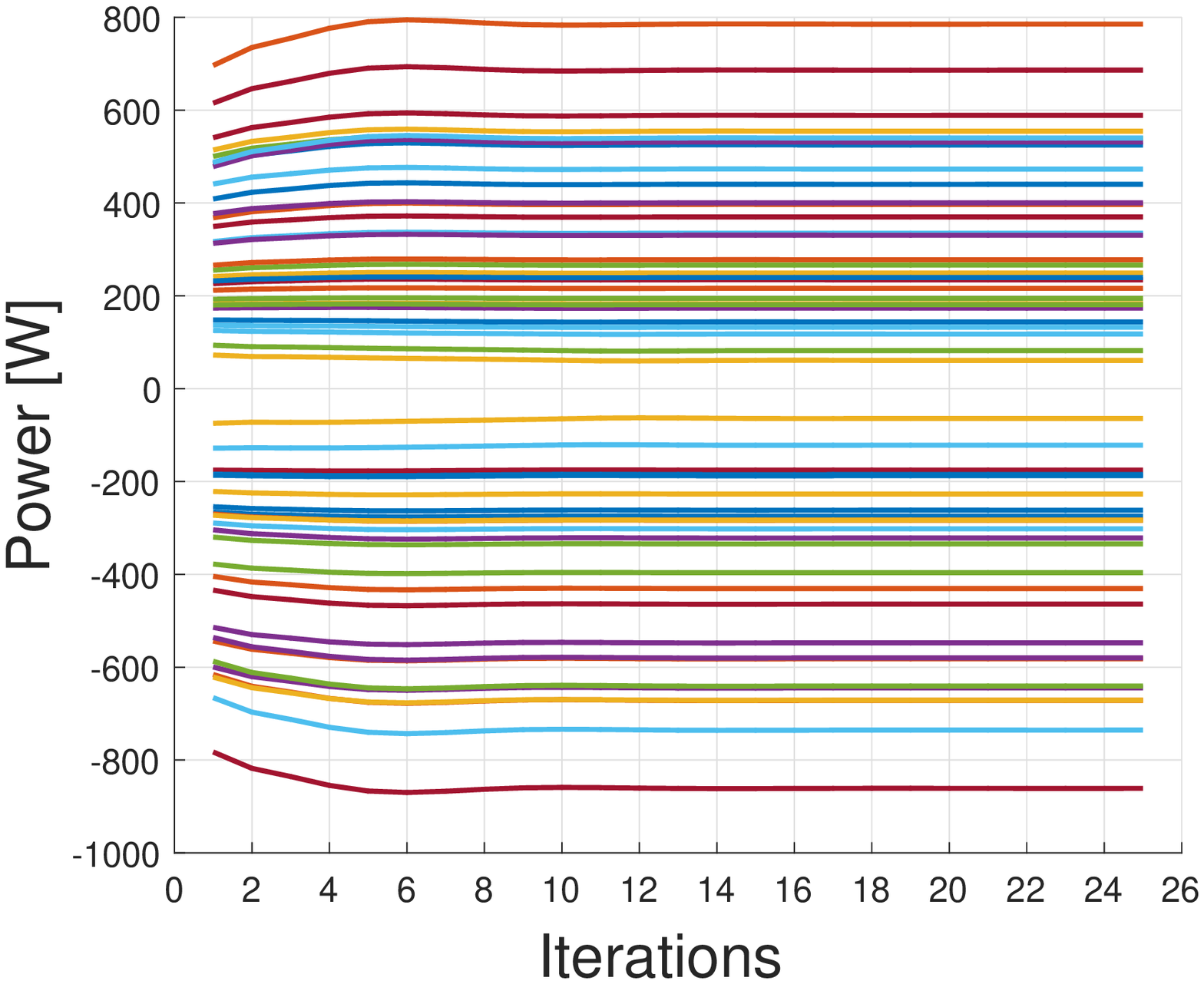}
		%%\vspace{-15mm}
		%\caption{Total traded power of all prosumers.}
		%\label{eu_power_1}
	%\end{figure}	
	%
	%\begin{figure}[htpb!]
		%\centering
		\includegraphics[scale=0.3]{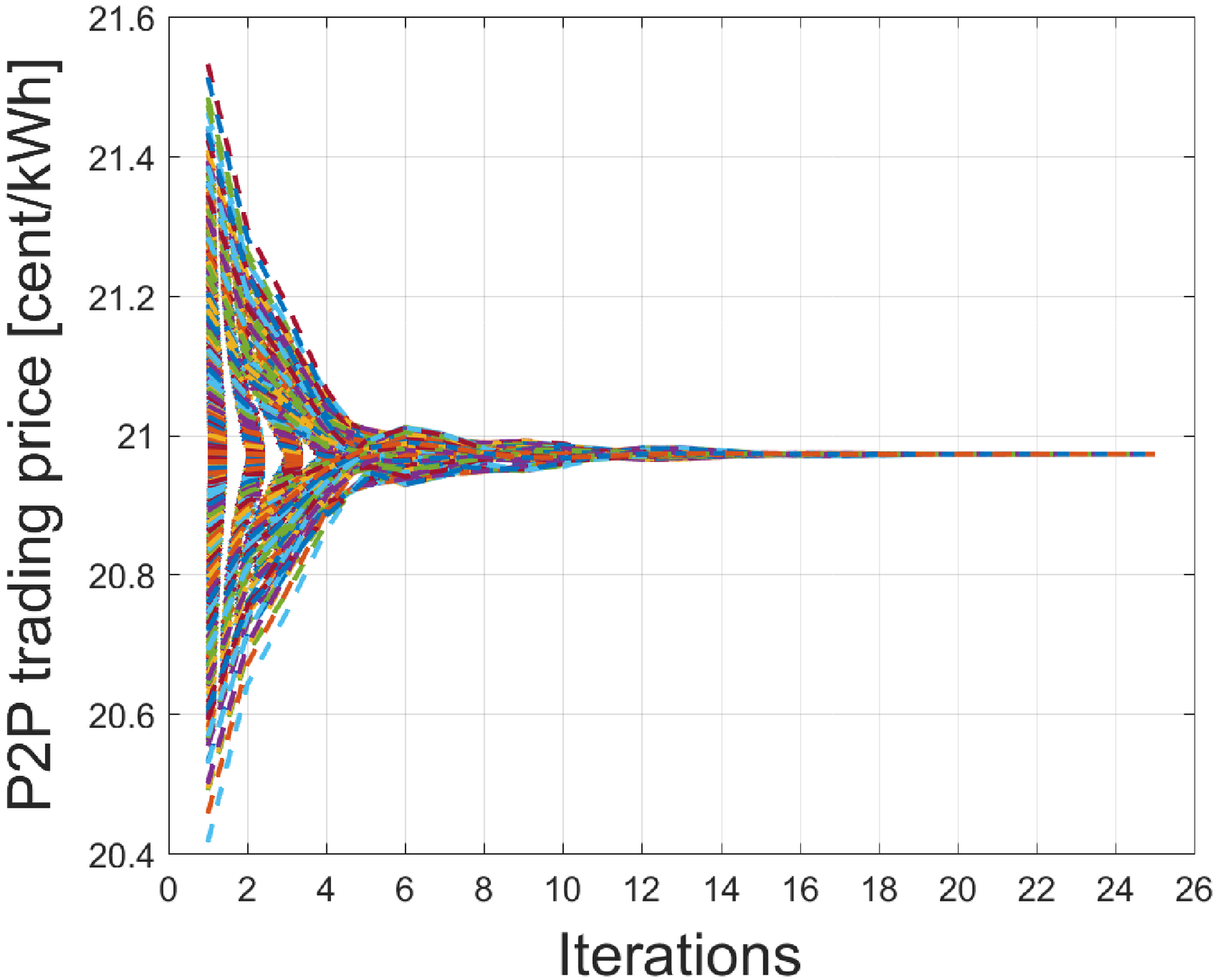}
		%\vspace{-15mm}
		\caption{Convergence of total traded power and optimal price for P2P trading between prosumers.}
		\label{eu_price_1}
	\end{figure}

As seen in Figure \ref{eu_1h_loads_solar}, at noon solar generation is maximum, while 25 solar-equipped nodes have maximum demands of 3.5kW, therefore they have at least 2kW solar power redundant which will be sold to other 30 nodes. As such, there are 25 selling and 30 buying prosumers, and %each seller is communicated with each buyer, hence 
the inter-prosumer communication graph has 750 edges.  
Initially, parameters $a_i$ and $b_i$ are randomly generated in the intervals $[0.005,0.009]$ and $[12.4,31.2]$, respectively. 

	\begin{figure}[htpb!]
		\centering
		\includegraphics[scale=0.3]{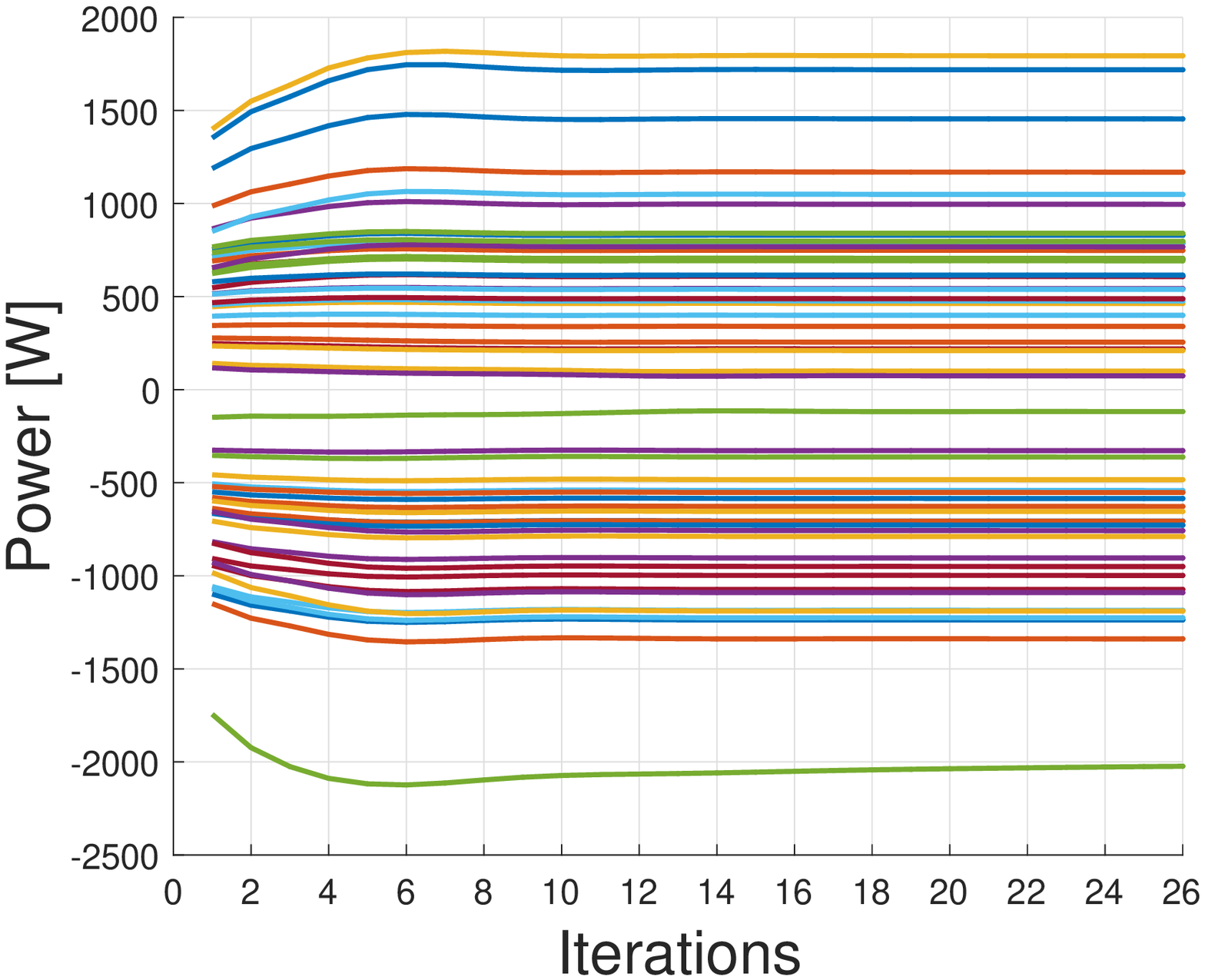}
		%%\vspace{-15mm}
		%\caption{Increased total traded power of all prosumers when $a_i$ are decreased.}
		%\label{eu_power}
	%\end{figure}	
	%
	%\begin{figure}[htpb!]
		%\centering
		\includegraphics[scale=0.3]{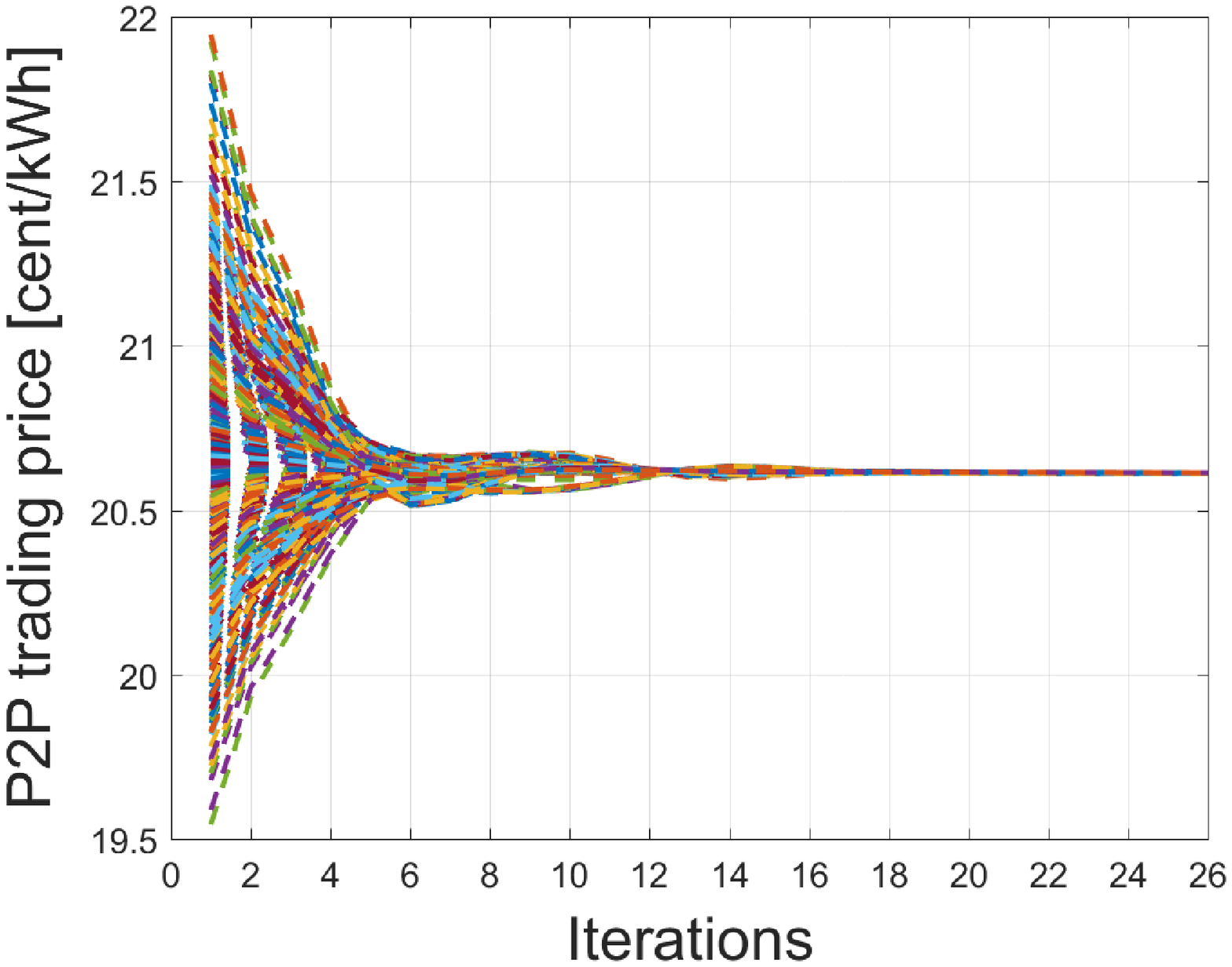}
		%\vspace{-15mm}
		\caption{Convergence of total traded power and optimal price for prosumers P2P trading when $a_i$ are decreased.}
		\label{eu_price}
	\end{figure}	

Running the proposed decentralized ADMM approach for P2P energy trading between prosumers gives us the results in Figure \ref{eu_price_1}. %The solving time for this case is 14.2s. 
As seen in Figure \ref{eu_price_1}, the total selling and buying powers of prosumers are far from their maximum capacities. Thus, using the proposed decentralized learning method for tuning prosumers parameters, all prosumers reduces their parameters $a_i$ by randomly regenerating them in the interval  $[0.002,0.006]$. The simulation results for the new values of $a_i$ are depicted in Figure \ref{eu_price}. It is obviously observed in Figure \ref{eu_price} that the total traded power of prosumers are significantly increased in absolute values, which validates the proposed learning strategy. The optimal prices are not much different between two cases. Similar simulation results are obtained at other time steps.

The merits of P2P energy trading here are as follows. First, it reduces power flows from and to the bulk grid by local consumption, hence eases power losses and voltage/frequency problems. Second, households with renewable generation can save their investment costs by selling excessive renewable energy to other neighboring households instead of storing with battery storage systems. 

%===============================================
\subsection{Scalability of The Proposed Approach}	

	\begin{figure}[htpb!]
		\centering
		\includegraphics[scale=0.3]{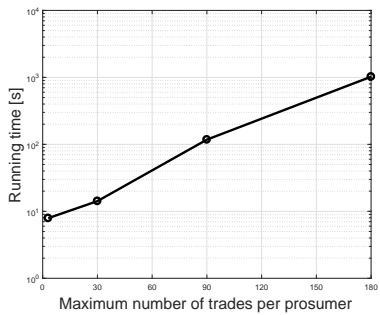}
		%\vspace{-15mm}
		\caption{Computational time of the proposed distributed P2P ADMM approach vs. system size.}	
		\label{admm_runtime}
	\end{figure}	
	
The proposed decentralized ADMM algorithm running times for the above two cases of IEEE European Low Voltage Test Feeder are 14.2s and 21.2s, which are just double of that for the synthetic system (c.f. Table \ref{table-1}), while the  maximum number of communicated peers (i.e. trades) per prosumer is 30, i.e. 10 times bigger. Next, the proposed algorithm is tested for larger systems, one with 75 selling and 90 buying prosumers, and the other with 150 selling and 180 buying prosumers. Running time for the former is 117.6s, while that for the latter is 1022.9s. As seen in Figure \ref{admm_runtime}, computational time of the proposed P2P ADMM approach is not exponentially increased with system size, hence is scalable well.

Note that the computational time shown in Figure \ref{admm_runtime} does not take into account the inter-peer communication time which may contain delays. In practical implementation, communication latency is an issue needs to be carefully handled for both centralized, distributed, and decentralized approaches, due to the tradeoff between communication load and computational ability. Centralized and distributed approaches have the advantages of stronger computational capability at the central unit and less total number of communication links over decentralized approaches. On the other hand, all communication tasks occur at the central unit in centralized and distributed optimization approaches, of which communication latency could become a critical issue, while communication load at each agent in decentralized approaches could be much fewer. However, communication delay is out of scope of the current research, and will be addressed in the future work.

%%%%%%%%%%%%%%%%%%%%%%%%%%%%%%%%%%%%%%%%%%%%%%%%%%%%%%%%%%%%%%%%%%%%%%%%%%%%%%%%%%%%%%%%%%%%%%%%%%%%%%%%%%%%%%%%%%%%
\section{Conclusion}
\label{sum}

This paper has proposed a  decentralized ADMM optimization approach for P2P energy markets and decentralized learning strategies for prosumers to obtain successful transactions and total traded powers as they expect. Analytical formulas for the amount of power and its associated transaction price in each trading between a pair of prosumers were derived, which reveal insights on the relation between individual trading with the total traded amount of each prosumer, trade weights, and prosumer cost function parameters. These serve as bases for the proposed decentralized learning strategies to tune prosumers cost function parameters, and for the choice of trade weights to attain strategic objectives such as pollutant emissions reduction.  The effectiveness and scalability of the proposed approaches were illustrated through different case studies.

In the next research, more complex models of P2P market participants, power flow constraints in the P2P energy systems, and the interaction of P2P markets with other markets should be investigated. Moreover, different factors such as communication delays, systematic way to choose trade weights, etc., should also be studied.

%%%%%%%%%%%%%%%%%%%%%%%%%%%%%%%%%%%%%%%%%%%%%%%%%%%%%%%%%%%%%%%%%%%%%%%%%%%%%%%%%%%%%%%%%%%%%%%%%%%%%%%%%%%%%%%%%%%%

\section*{Appendix}

%-------------------------------------------------------
\subsection{Proof of Theorem \ref{opt-char}}
%\label{apdix-1}

When $d_{ij}=0 ~\forall \, i,j=1,\ldots,n$, and the communication graph between successfully traded peers is connected,  it is easy to see from \eqref{opt-Lagr-mtl} that $\lambda_{ij}^{\ast}$ are the same for all $j \in \tN_i$, hence are the same for all $i=1,\ldots,n$ due to the symmetry of the trading and the connectedness of the communication graph. The unique price is calculated by $2a_i P_{i,tr}^{\ast} + b_i = \lambda^{\ast}$. Based on this, we can easily derive $\lambda^{\ast}$ as in \eqref{eq-price} due to the fact that $\sum_{i=1}^{n} P_{i,tr}^{\ast} = 0$, and then obtain $P_{i,tr}^{\ast}$ as in \eqref{eq-power}. 

Next, utilizing the KKT conditions to the social welfare maximization problem \eqref{swm}, we also obtain the clearing energy price equals to $\left. \frac{\partial C_i}{\partial P_{i,tr}} \right|_{P_{i,tr}^{\ast}}  = 2a_i P_{i,tr}^{\ast} + b_i$. As such, the energy price in this market is the same with that in P2P market without trade weights. Consequently, it is obvious that the total traded energy of each peer in the P2P market is equal to that in the pool-based market. 
In case this graph is unconnected, i.e. the considering P2P market is clustered into smaller P2P markets, each cluster inherits the above properties. 

Finally, if $d_{ij} \neq 0$, then \eqref{opt-Lagr-mtl} can be stacked together to obtain $E^T P_{tr}^{\ast} = vec(b_j + d_{ji} - b_i - d_{ij})$. In addition, the condition $\sum_{i=1}^{n} P_{i,tr}^{\ast} = 0$ is rewritten as $\alpha^T P_{tr}^{\ast}=0$. Therefore, we have \eqref{eq-power-1}. 

%-------------------------------------------------------
\subsection{Proof of Theorem \ref{opt-char-admm}}
Substituting \eqref{price-constraint} into \eqref{price-eq} gives us
\begin{align*}
%\label{price-eq}
& 2a_iP_{i,tr}^{k+1} + (\rho+\phi)P_{ij}^{k+1} + v_{ij}^{k}  \\
&= 2a_jP_{j,tr}^{k+1} - (\rho+\phi)P_{ij}^{k+1} + v_{ji}^{k}
\end{align*}
which is \eqref{opt-var-1}. Using this to compute $\lambda_{ij}^{k+1}$ leads to \eqref{opt-var-11}. 
Then summing up \eqref{opt-var-1} for all $j \in \tN_i$ gives us
{\small
\begin{align*}
P_{i,tr}^{k+1}  
= \frac{\displaystyle \sum_{j \in \tN_i} v_{ji}^{k+1} + \sum_{j \in \tN_i} 2a_jP_{j,tr}^{k+1} - \sum_{j \in \tN_i} v_{ij}^{k+1} - 2n_ia_iP_{i,tr}^{k+1}}{2(\rho+\phi)} 
\end{align*}
which is equivalent to
\begin{align}
\label{opt-var-2}
& 2(\rho+\phi+n_ia_i)P_{i,tr}^{k+1} - \sum_{j \in \tN_i} 2a_jP_{j,tr}^{k+1} \notag \\
&= \sum_{j \in \tN_i} v_{ji}^{k+1}  - \sum_{j \in \tN_i} v_{ij}^{k+1}  
\end{align}
Equation \eqref{opt-var-2} is rewritten exactly as \eqref{Ptr}.}  
Next, the convergence of the proposed ADMM algorithm follows that provided in \cite{Deng2017}, hence we omit the proof here for brevity. 
Lastly, substituting the converged solutions to \eqref{price-eq} leads to $\lambda_{ij}^{\ast}  = 2a_iP_{i,tr}^{\ast} + b_i + d_{ij} + \rho u_{ij}^{\ast}$. Thus, as $\rho \rightarrow 0$, $\lambda_{ij}^{\ast}  = 2a_iP_{i,tr}^{\ast} + b_i + d_{ij}$, solutions in Theorem \ref{opt-char}.

%%%%%%%%%%%%%%%%%%%%%%%%%%%%%%%%%%%%%%%%%%%%%%%%%%%%%%%%%%%%%%%%%%%%%%%%%%%%%%%%%%%%%%%%%%%%%%%%%%%%%%%%%%%%%%%%%%%
\bibliographystyle{plain}
\bibliography{dp}
%\vspace{-1.2cm}

\end{document}